\documentclass[11pt,a4paper]{article}
\pdfoutput=1
\usepackage{jheppub}
\usepackage{rotating}
\usepackage{array}
\usepackage{amsmath}
\usepackage[normalem]{ulem}
\usepackage{slashed}
\preprint{DESY 15-182}

\title{Implications of unitarity and gauge invariance for simplified dark
  matter models}

\author[a]{Felix Kahlhoefer,}
\author[a]{Kai Schmidt-Hoberg,}
\author[b]{Thomas Schwetz}
\author[b,c]{and Stefan Vogl}

\affiliation[a]{DESY, Notkestra\ss e 85, D-22607 Hamburg, Germany}
\affiliation[b]{Institut f\"ur Kernphysik, Karlsruher Institut f\"ur Technologie (KIT), D-76021 Karlsruhe, Germany}
\affiliation[c]{Oskar Klein Centre for Cosmoparticle Physics, Department of Physics, Stockholm University, SE-10691 Stockholm, Sweden}

\emailAdd{felix.kahlhoefer@desy.de}
\emailAdd{kai.schmidt.hoberg@desy.de}
\emailAdd{schwetz@kit.edu}
\emailAdd{stefan.vogl@fysik.su.se}

\abstract{We show that simplified models used to describe the interactions of dark matter with Standard Model particles do not in general respect gauge invariance and that perturbative unitarity may be violated in large regions of the parameter space. The modifications necessary to cure these inconsistencies may imply a much richer phenomenology and lead to stringent constraints on the model. We illustrate these observations by considering the simplified model of a fermionic dark matter particle and a vector mediator. Imposing gauge invariance then leads to strong constraints from dilepton resonance searches and electroweak precision tests. Furthermore, the new states required to restore perturbative unitarity can mix with Standard Model states and mediate interactions between the dark and the visible sector, leading to new experimental signatures such as invisible Higgs decays. The resulting constraints are typically stronger than the `classic' constraints on DM simplified models such as monojet searches and make it difficult to avoid thermal overproduction of dark matter.}


\begin{document}
\maketitle
\flushbottom

\section{Introduction}

After the successful discovery of a Higgs Boson consistent with the predictions of the Standard Model (SM), the focus of the current and upcoming runs of the Large Hadron Collider (LHC) at 13 TeV will be to discover evidence for physics beyond the SM. Among the prime targets of this search is dark matter (DM), which has so far only been observed via its gravitational interactions at astrophysical and cosmological scales. Since no particle within the SM has the required properties to explain these observations, DM searches at the LHC are necessarily searches for new particles. 

In fact, LHC DM searches are also likely to be searches for new interactions. Given the severe experimental constraints on the interactions between DM and SM particles, it is a plausible and intriguing possibility that the DM particle is part of a (potentially rich) hidden sector, which does not couple directly to SM particles or participate in the known gauge interactions. In this setup, the visible sector interacts with the hidden sector only via one or several new mediators, which have couplings to both sectors.

In the simplest case the mass of these mediators is large enough that they can be integrated out and interactions between DM particles and the SM can be described by higher-dimensional contact interactions~\cite{Beltran:2008xg,Beltran:2010ww}. This effective field theory (EFT) approach has been very popular for the analysis and interpretation of DM searches at the LHC~\cite{Goodman:2010ku,Fox:2011pm,Rajaraman:2011wf}. Nevertheless, as any effective theory it suffers from the problem that unitarity breaks down if the relevant energy scales become comparable to the cut-off scale of the theory~\cite{Shoemaker:2011vi,Fox:2012ee,Busoni:2013lha,Busoni:2014sya,Xiang:2015lfa} (for other examples of applying unitarity arguments in the DM context see refs.~\cite{Griest:1989wd, Walker:2013hka, Endo:2014mja, Hedri:2014mua}).

The easiest way to avoid this problem appears to be to explicitly retain the (lightest) mediator in the theory. The resulting models are referred to as simplified DM models, in which couplings are only specified after electroweak symmetry breaking (EWSB) and no ultraviolet (UV) completion is provided~\cite{Abdallah:2015ter}. Compared to the EFT approach, simplified models have a richer phenomenology~\cite{Busoni:2013lha,Buchmueller:2013dya,Buchmueller:2014yoa,Harris:2014hga,Garny:2014waa,Buckley:2014fba,Jacques:2015zha,Alves:2015dya,Choudhury:2015lha}, including explicit searches for the mediator itself~\cite{Frandsen:2012rk,Fairbairn:2014aqa,Chala:2015ama}. Moreover, it is possible to achieve the DM relic abundance in large regions of parameter space~\cite{Busoni:2014gta,Chala:2015ama,Blennow:2015gta}. Constraining the parameter space of simplified DM models is therefore a central objective of experimental collaborations~\cite{Khachatryan:2014rra, Aad:2015zva,Abercrombie:2015wmb}.

In the present work we focus on the case of a spin-1 $s$-channel
mediator~\cite{Dudas:2009uq,Fox:2011qd,Frandsen:2012rk,Alves:2013tqa,Arcadi:2013qia,Jackson:2013pjq,Jackson:2013rqp,Duerr:2013lka,Duerr:2014wra,Lebedev:2014bba,Hooper:2014fda,Martin-Lozano:2015vva,Alves:2015pea,Alves:2015mua,Blennow:2015gta,Duerr:2015wfa,Heisig:2015ira}. Our central observation is that the simplified model approach is not generally
sufficient to avoid the problem of unitarity violation at high
energies and that further amendments are required if the model
  is to be both simple and realistic. In particular, a spin-1
mediator with axial couplings violates perturbative unitarity at large
energies, pointing towards the presence of additional new physics to
restore unitarity. 

Indeed, the simplest way to restore unitarity is to assume that the
spin-1 mediator is the gauge boson of an additional $U(1)'$ gauge
symmetry~\cite{Holdom:1985ag,Babu:1997st} and that its mass as well as
the DM mass are generated by a new Higgs field in the hidden
sector. The famous Lee-Quigg-Thacker bound~\cite{Lee:1977eg} implies
that the additional Higgs boson cannot be arbitrarily heavy and may
therefore play an important role for LHC and DM phenomenology. In particular,
it can mix with the SM-like Higgs boson and mediate interactions
between DM particles and quarks.

Furthermore, we require for a consistent simplified DM model that the coupling structure respects gauge invariance of the full SM gauge group before EWSB (see~\cite{Bell:2015sza} for a similar discussion in the EFT context). If the mediator has axial couplings to quarks, this requirement implies that the new mediator will also have couplings to leptons and mixing with the SM $Z$ boson, both of which are tightly constrained by experiments. Much weaker constraints are obtained for the simplified DM model containing a spin-1 mediator with vectorial couplings to quarks.  Constraints from direct detection can be evaded if the mediator has only axial couplings to DM, which naturally arises in the case that the DM particle is a Majorana fermion. We discuss the importance of loop-induced mixing effects in this context, which can play a crucial role for both direct detection experiments and LHC phenomenology.

The outline of the paper is as follows. Starting from a simplified model for a \mbox{spin-1} \mbox{$s$-channel} mediator, we explore in section~\ref{sec:unitarity} the implications of perturbative unitarity, deriving a number of constraints on the model parameters and in particular an upper bound on the scale of additional new physics. In section~\ref{sec:higgs} we then consider the case where this additional new physics is a Higgs field in the hidden sector and derive an upper bound on the mass of the extra Higgs boson. We then discuss additional constraints on the SM couplings implied by gauge invariance. Section~\ref{sec:axial} focuses on the case of non-zero axial couplings between SM fermions and the mediator, whereas in section~\ref{sec:vector} we assume that the SM couplings of the mediator are purely vectorial. 
Finally, we discuss the experimental implications of a possible mixing between the SM Higgs and the hidden sector Higgs in section~\ref{sec:higgsmixing}. A discussion of our results and our conclusions are presented in section~\ref{sec:discussion}.

\section{Unitarity constraints on simplified models}
\label{sec:unitarity}

\subsection{Brief review of $S$ matrix unitarity constraints}

Consider the scattering matrix element $\mathcal{M}_{if}(s, \cos
\theta)$ between 2-particle initial and final states ($i,f$), with $\sqrt{s}$
and $\theta$ being the centre of mass energy and scattering angle,
respectively. We define the helicity matrix element for the $J$th
partial wave by
\begin{equation}\label{eq:Jexpansion}
  \mathcal{M}_{if}^J(s) = \frac{1}{32\pi} \beta_{if} 
  \int_{-1}^1 \mathrm{d}\cos \theta \, d^J_{\mu \mu'}(\theta) \, \mathcal{M}_{if}(s, \cos \theta) \,,
\end{equation}
where $d^J_{\mu \mu'}$ is the $J$th Wigner d-function, $\mu$ and $\mu'$ denote the total spin of the initial and the final state (see e.g.~\cite{Chanowitz:1978mv}), and $\beta_{if}$ is a
kinematical factor. In the high-energy limit $s
\to \infty$, which we are going to consider below, $\beta_{if} \to 1$. The right-hand side of eq.~\eqref{eq:Jexpansion} is to be multiplied with a factor of $1/\sqrt{2}$ each if the initial or final state particles are identical~\cite{Schuessler:2007av}.
Unitarity of the $S$ matrix implies
\begin{align}
  {\rm Im}(\mathcal{M}_{ii}^J) & = \sum_f | \mathcal{M}_{if}^J|^2 \nonumber\\
  &=
  | \mathcal{M}_{ii}^J|^2 + \sum_{f \neq i} | \mathcal{M}_{if}^J|^2 \ge | \mathcal{M}_{ii}^J|^2
  \label{eq:unity}
\end{align}
for all $J$ and all $s$. The sum over $f$ in the first line runs over
all possible final states. Restricting these to be all possible
2-particle states leads to a conservative bound. 
If the relation \eqref{eq:unity} is strongly violated
for matrix elements calculated at leading order in perturbation theory
one can conclude that either higher-order terms in perturbation theory
restore unitarity (i.e.\ break-down of
perturbativity) or that the theory is not complete and additional
contributions to the matrix element are needed. 

From eq.~\eqref{eq:unity} one obtains the necessary conditions
\begin{equation}\label{eq:unitary-bound}
  0 \le {\rm Im}({\mathcal{M}}_{ii}^J) \le 1\,, \quad
  \left| \text{Re}({\mathcal{M}}_{ii}^J) \right| \le \frac{1}{2} \,.
\end{equation}
In the following we will apply these inequalities to leading-order matrix elements in order to identify regions in parameter space where perturbative unitarity is violated. Since these matrix elements are always real in the present context, only the second constraint will be relevant.

If the matrix $\mathcal{M}_{if}^J$ is diagonalized the
inequality in eq.~\eqref{eq:unity} becomes an equality. Hence,
stronger constraints can be obtained by considering the full
transition matrix connecting all possible 2-particle states with each
other (or some submatrix thereof) and calculating the eigenvalues of
that matrix. Then the bounds from eq.~\eqref{eq:unitary-bound} have to
hold for each of the eigenvalues~\cite{Schuessler:2007av}.

We note that $d^J_{00}(\theta) = P_J(\cos \theta)$, where $P_J$ are the Legendre polynomials. If initial and final state both have zero total spin, eq.~\eqref{eq:Jexpansion} therefore becomes identical to the familiar partial wave expansion of the matrix element. In the following we will focus on the $J=0$ partial wave, which typically provides the strongest constraint. Since $d^0_{\mu\mu'}$ is non-zero only for $\mu = \mu' = 0$, we then obtain from eq.~\eqref{eq:Jexpansion}
\begin{equation}
  \mathcal{M}_{if}^0(s) = \frac{1}{64\pi} \beta_{if} \, \delta_{\mu0} \delta_{\mu'0} 
  \int_{-1}^1 \mathrm{d}\cos \theta \, \mathcal{M}_{if}(s, \cos \theta) \,.
\end{equation}

\subsection{Application to a simplified model with a $Z'$ mediator}

Let us consider a simplified model for a spin-1 mediator $Z'^\mu$ with mass $m_{Z'}$ and a Dirac DM particle $\psi$ with mass $m_\text{DM}$.\footnote{In the case of Majorana DM the vector current vanishes and hence there can only be an axial coupling on the DM side. We will come back to this case shortly but will consider Dirac DM here to allow for both vectorial and axial couplings.} The most general coupling structure is captured by the following Lagrangian:
\begin{equation}\label{eq:L_VA}
\mathcal{L} =  - \sum_{f = q,l,\nu} Z'^\mu \, \bar{f} \left[ g_{f}^V \gamma_\mu + g_f^A \gamma_\mu \gamma^5 \right] f - Z'^\mu \, \bar{\psi} \left[ g_\text{DM}^V \gamma_\mu + g_\text{DM}^A \gamma_\mu \gamma^5 \right] \psi \; .\\
\end{equation}
Although these interactions appear renormalisable, the presence of a massive vector boson implies that perturbative unitarity may be violated at large energies. In the following, we will study this issue in detail and derive constraints on the parameter space of the model.

Let us first consider diagrams between 2-fermion states with the $Z'$
as mediator. The appropriate propagator for the mediator is
\begin{equation}
\langle Z'^\mu(k)Z'^\nu(-k)\rangle =  \frac{1}{k^2 - m_{Z'}^2} \left(g^{\mu\nu} - \frac{k^\mu k^\nu}{m_{Z'}^2}\right) \;,
\end{equation}
where $k^\mu$ is the momentum of the mediator.  For the case of a
gauge boson this corresponds to unitary gauge in which the Goldstone
boson has been absorbed. Since we are interested in the high-energy
behaviour of the theory we concentrate on the second term, which does
not vanish in the limit $k \rightarrow \infty$. This corresponds to
restricting to the longitudinal component of the mediator, $Z'_L$,
which dominates at high energy \cite{Chanowitz:1978mv}.\footnote{It
  turns out that for certain processes the transversal part of the
  propagator leads to a logarithmic divergence for $m_{Z'}^2 \ll s$.
  This divergence is not related to the UV completeness of the theory,
  but signals breakdown of perturbativity in the IR, see also
  \cite{Hedri:2014mua}. By restricting to the longitudinal components
  of the $Z'$ \cite{Chanowitz:1978mv} we can avoid the occurence of
  those IR divergences.}  For
instance, considering DM annihilations, we can contract the longitudinal
part of the propagator with the DM current.  Making use of $k = p_1 +
p_2$, where $p_1$ and $p_2$ are the momenta of the two DM particles in
the initial state, leads to a factor
\begin{align}
 k^\mu \bar{v}(p_2) \left( g^V_\text{DM} \gamma_\mu + g^A_\text{DM} \gamma_\mu \gamma^5 \right) u(p_1)  
 & = \bar{v}(p_2) \left[ g^V_\text{DM} (\slashed{p}_2 + \slashed{p}_1) + g^A_\text{DM} (\slashed{p}_2 \gamma^5 - \gamma^5 \slashed{p}_1) \right] u(p_1) \nonumber \\
 & = - 2 \, g^A_\text{DM} \, m_\text{DM} \, \bar{v}(p_2) \gamma^5 u(p_1) \;.
\end{align}
Hence, the second term in the propagator behaves exactly like
a pseudoscalar with mass $m_{Z'}$ and couplings to DM equal to $2 \,
g^A_\text{DM} \, m_\text{DM} / m_{Z'}$, just like the Goldstone boson
present in Feynman gauge. Note that the term is independent of the vector couplings. The same argument holds for the quark
couplings, which are found to be given by $2 \, g^A_f \, m_f /
m_{Z'}$.
This consideration suggests that perturbative unitarity will not only lead to bounds on $g^{V,A}$, but also on the combination $g^A_f \, m_f / m_{Z'}$.

We can make this statement more precise by applying the methods outlined in the previous subsection to the self-scattering of two DM particles or two SM fermions.
We obtain for any fermion $f$ with axial couplings $g^A_f \neq 0$ that the fermion mass must satisfy the bound
\begin{equation}
 m_f \lesssim \sqrt{\frac{\pi}{2}} \frac{m_{Z'}}{g^A_f} \; .
 \label{eq:DMmass}
\end{equation}
Here $f$ can be any fermion, including SM fermions and the DM
particle. As suggested by the above discussion we do not obtain any
bound on the masses of fermions with purely vectorial couplings, nor
on the scale of new physics.

Let us now turn to the discussion of processes involving $Z'$ in the
external state, in particular $Z'$ with longitudinal polarisation. For
concreteness, we study the process $\psi \bar{\psi} \rightarrow Z'_L
Z'_L$.\footnote{Note that this process corresponds to an off-diagonal
  element of $\mathcal{M}_{if}$, with $i \neq f$, whereas the bounds
  from eq.~\eqref{eq:unitary-bound} apply for diagonal elements. In
  order to apply the unitarity constraint we consider the $2\times 2$
  submatrix of $\mathcal{M}_{if}$ spanned by the states $\psi
  \bar{\psi}$ and $Z'_L Z'_L$. For $s\to \infty$ only the off-diagonal
  element survives, and hence the eigenvalues of the matrix become
  equal to the off-diagonal element, and we can apply
  eq.~\eqref{eq:unitary-bound}.} At large momenta, $k^2 \gg
m_{Z'}^2$, the polarisation vectors of the gauge bosons can be
replaced by $\epsilon_L^\mu(k) = k^\mu / m_{Z'}$. One might therefore
expect the matrix element for this process to grow proportional to $s
/ m_{Z'}^2$. However, such a term is absent due to a cancellation
between the $t$- and $u$-channel diagram. To obtain a non-zero
contribution, one needs to include a mass insertion along the fermion
line~\cite{Shu:2007wg}. It turns out that the contribution
proportional to $g^V_\text{DM}$ still cancels in this case and that
the leading contribution at high energies becomes proportional to
$(g^A_\text{DM})^2 \sqrt{s} \, m_\text{DM} / m_{Z'}^2$. As a result,
perturbative unitarity is violated
unless~\cite{Hosch:1996wu,Shu:2007wg,Babu:2011sd}\footnote{Our result differs from the one in~\cite{Shu:2007wg} by a factor $1/\sqrt{2}$.}
\begin{equation}\label{eq:s}
\sqrt{s} < \frac{\pi \, m_{Z'}^2}{(g^A_\text{DM})^2 \, m_\text{DM}} \; .
\end{equation}
For larger energies new physics must appear to restore unitarity.  This can be accomplished by including an additional diagram with an $s$-channel Higgs boson, since both contributions have the same high-energy behaviour. The consideration above implies an upper bound on the mass of the Higgs that breaks the $U(1)'$ and gives mass to the $Z'$:
\begin{equation}
m_s < \frac{\pi \, m_{Z'}^2}{(g^A_\text{DM})^2 \, m_\text{DM}} \; .
\label{eq:higgsmass}
\end{equation}
We will discus the consequences  of such an extension of the minimal model in section~\ref{sec:higgs}.

\begin{figure}[tb]
\centering
\includegraphics[height=0.3\textheight]{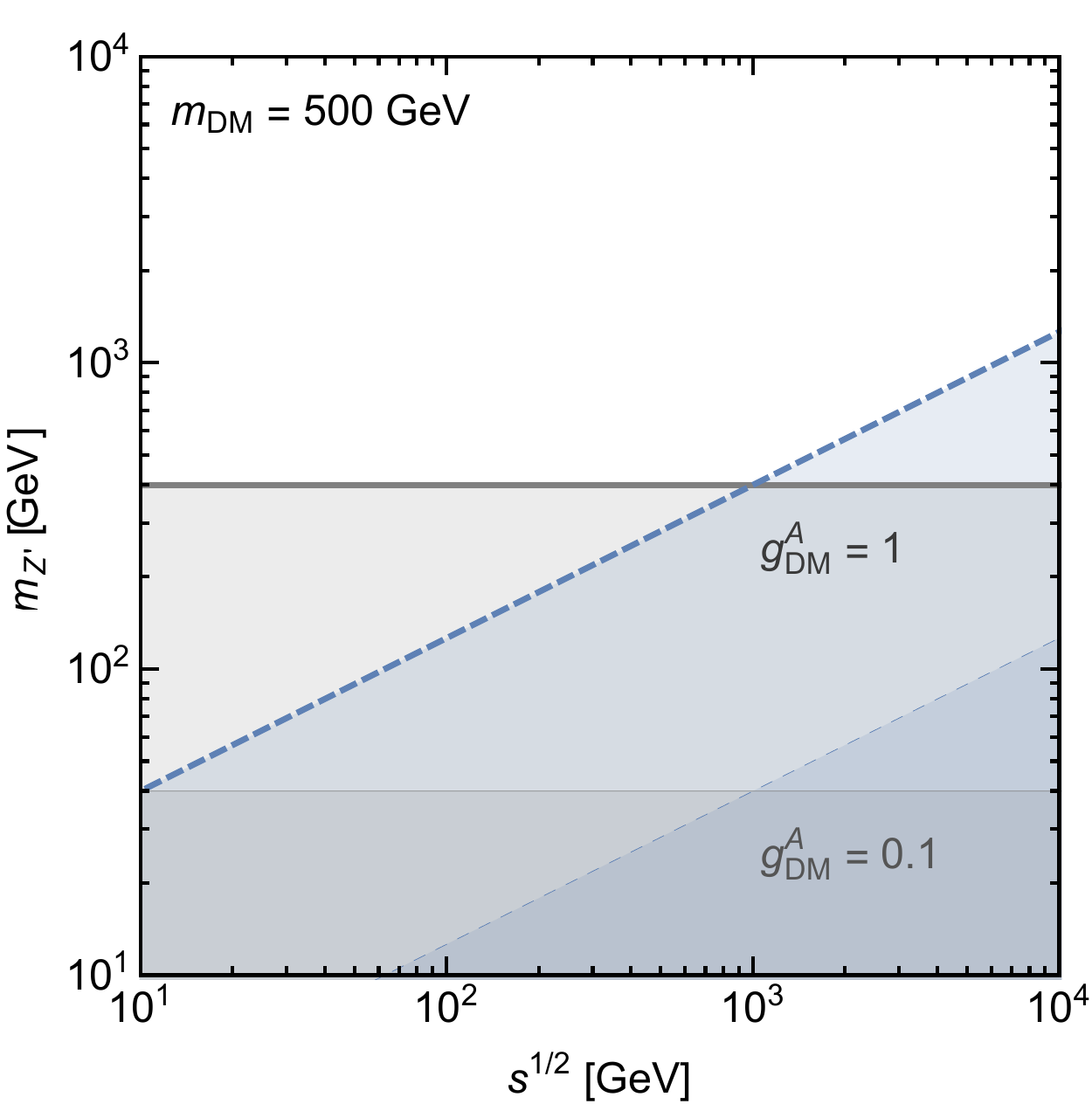}
\caption{Parameter space forbidden by the requirement of perturbative unitarity in the $\sqrt{s}-m_{Z'}$ plane for $m_\text{DM}=500\:\text{GeV}$. The constraint resulting from DM scattering is shown in grey (solid), the constraint resulting from DM annihilation into $Z'$s is shown in blue (dashed). Thick (thin) lines correspond to $g^A_\text{DM}=1$ ($g^A_\text{DM}=0.1$). In these cases, the $Z'$ can never be lighter than about $400\:\text{GeV}$ ($40\:\text{GeV}$) irrespective of the UV completion.}
\label{fig:ubound}
\end{figure}

In summary we have found that there are two different types of
constraints on the parameters of this simplified model, even for
perturbative couplings. For non-vanishing axial couplings there is an
energy scale for which the theory violates perturbative unitarity and
needs to be UV completed, see eq.~\eqref{eq:s}.  In addition, imposing that the coupling between the longitudinal component of the vector mediator and the DM particle remain perturbative, we find that the
vector mediator cannot be much lighter than the DM, see eq.~\eqref{eq:DMmass}. This constraint is not related to missing degrees of freedom and is therefore completely
independent of the UV completion. We illustrate both constraints in figure~\ref{fig:ubound} for different axial couplings
and a DM mass $m_\text{DM} = 500\:\text{GeV}$.

\begin{figure}[tb]
\centering
\includegraphics[height=0.3\textheight]{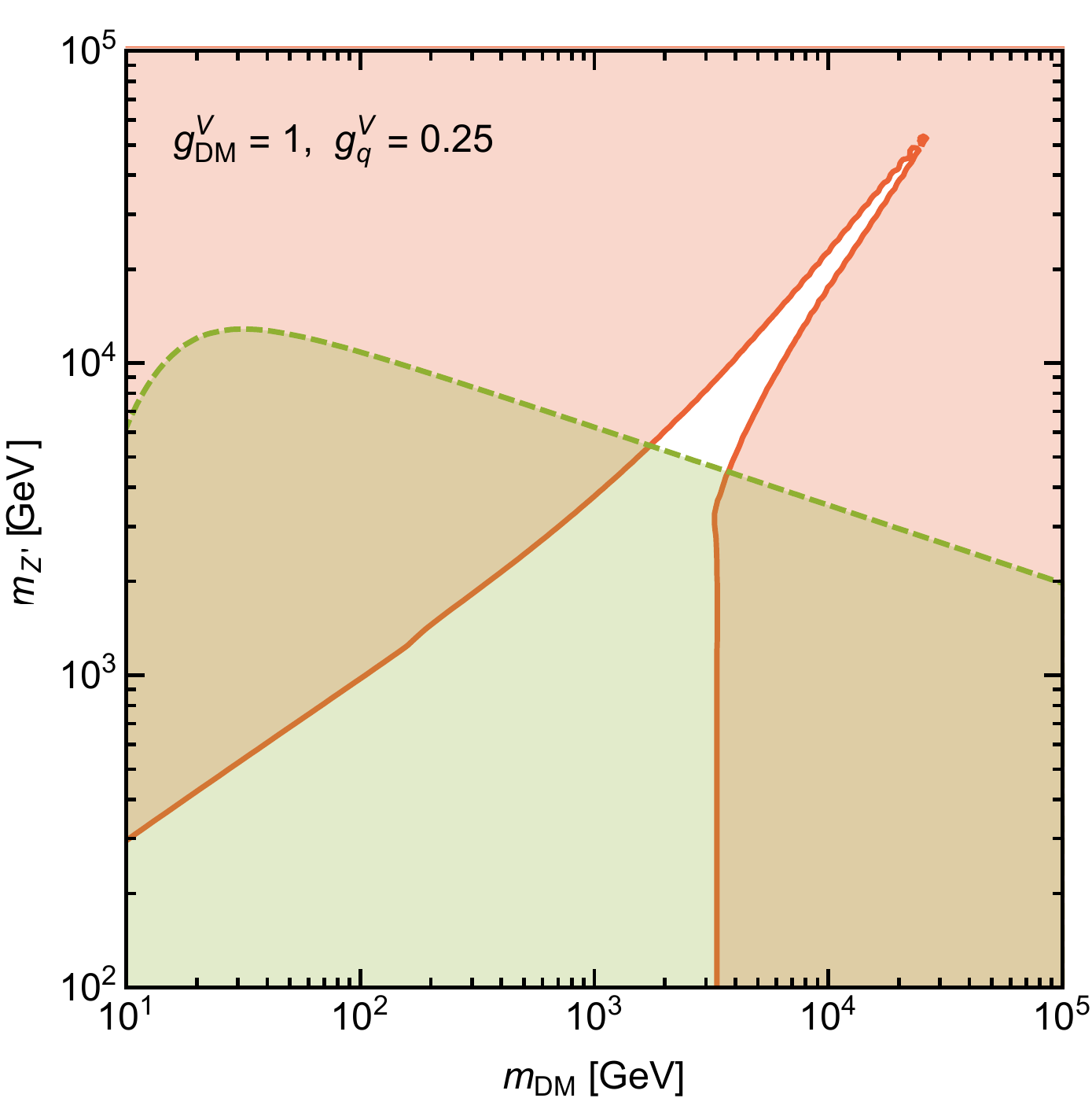}
\caption{Vector(SM)--Vector(DM): Parameter space excluded by the bound on the spin-independent DM-nucleon scattering cross section from LUX (green, dashed) and the parameter region where the DM annihilation cross section becomes so small that DM is overproduced in the early Universe (red, solid).}
\label{fig:vector}
\end{figure}

To conclude this section, we emphasise that for pure vector couplings of the $Z'$ ($g^A_\text{DM} = g^A_f = 0$) the simplified model considered in this section is well-behaved in the UV in the sense that there is no problem with perturbative unitarity.\footnote{As discussed below there can be anomalies which require additional fermions.} Indeed in this specific case a bare mass term for the dark matter is allowed such that it is sufficient to generate the vector boson mass via a Stueckelberg mechanism without the need for additional degrees of freedom~\cite{Stueckelberg:1900zz,Kors:2005uz}. However, this specific coupling configuration is highly constrained, since it is very difficult to evade bounds from direct detection experiments and still reproduce the observed DM relic abundance. This is illustrated in figure~\ref{fig:vector} where we show the parameter region excluded by the bound on the spin-independent DM-nucleon scattering cross section from LUX~\cite{Akerib:2013tjd} and the parameter region where the DM annihilation cross section becomes so small that DM is overproduced in the early Universe. One can clearly see that only a finely-tuned region of parameter space close to the resonance $m_\text{DM} = m_{Z'}/2$ is still allowed. For the rest of the paper, we will therefore not consider this case further and always assume that at least one of the vector couplings vanishes such that direct detection constraints can be weakened.

\section{Including an additional Higgs field}
\label{sec:higgs}

As we have seen in the previous section, for non-zero axial couplings the simplified model violates perturbative unitarity at high energies, implying that additional new physics must appear below these scales. This observation motivates a detailed discussion of how to generate the vector boson mass from an additional Higgs mechanism. To restore unitarity let us therefore now consider the case that the $Z'$ is the gauge boson of a new $U(1)'$ gauge group. To break this gauge group and give a mass to the $Z'$, we introduce a dark Higgs singlet $S$, which needs to be complex in order to allow for a $U(1)'$ charge. We then obtain the following Lagrangian
\begin{equation}
 \mathcal{L} = \mathcal{L}_\text{SM}  + \mathcal{L}_\text{DM} + \mathcal{L}'_\text{SM} + \mathcal{L}_\text{S} \; ,
\end{equation}
where the first term is the usual SM Lagrangian and the second term describes the interactions of DM. The third term contains the interactions between SM states and the new $Z'$ gauge boson while the fourth term contains the extended Higgs sector. 

\subsection{Implications for the dark sector}

As mentioned above, it is well-motivated from a phenomenological perspective to consider the case that vector couplings to the $Z'$ mediator vanish in at least one of the two sectors, so that direct detection is suppressed. On the DM side this is naturally achieved for a Majorana fermion, which we will focus on from now. We therefore write
\begin{equation}
\psi =
\left(
    \begin{array}{c}
      \chi \\
      \epsilon \chi^\ast
    \end{array}
  \right) \; ,
\end{equation}
where $\chi$ is a Weyl spinor. We assume that $\chi$ carries a charge $q_\text{DM}$ under the new $U(1)'$ gauge group, such that under a gauge transformation
\begin{equation}
\psi \rightarrow \exp\left[i \, g' q_\text{DM} \, \alpha(x) \, \gamma^5\right] \psi \; ,
\end{equation}
where $g'$ is the gauge coupling of the new $U(1)'$.
The kinetic term for $\psi$ can hence be written as
\begin{equation}
\mathcal{L}_\text{kin} = \frac{1}{2} \bar{\psi} (i \slashed{\partial} - g' \, q_\text{DM} \, \gamma^5 \slashed{Z}') \psi = \frac{i}{2} \bar{\psi} \slashed{\partial} \psi - \frac{1}{2} g_\text{DM}^A Z'^\mu \bar{\psi} \gamma^5 \gamma_\mu \psi \; ,
\end{equation}
with $g_\text{DM}^A \equiv g' q_\text{DM}$.
The $U(1)'$ charge forbids a Majorana mass term. Nevertheless, if the Higgs field $S$ carries charge $q_S = - 2 q_\text{DM}$, we can write down the gauge-invariant combination
\begin{equation}
\mathcal{L}_\text{mass} = -\frac{1}{2} y_\text{DM} \bar{\psi} (P_L S + P_R S^\ast) \psi \; .
\end{equation}
Including the kinetic and potential terms for the Higgs singlet, the
full dark Lagrangian therefore reads
\begin{align}
 \mathcal{L}_\text{DM} = & \frac{i}{2} \bar{\psi} \slashed{\partial} \psi - \frac{1}{2} g_\text{DM}^A Z'^\mu \bar{\psi} \gamma^5 \gamma_\mu \psi - \frac{1}{2} y_\text{DM} \bar{\psi} (P_L S + P_R S^\ast) \psi \,, \nonumber \\
\mathcal{L}_S = &  \left[ (\partial^\mu + i \, g_S \, Z'^\mu) S \right]^\dagger \left[ (\partial_\mu + i \, g_S \, Z'_\mu) S \right] + \mu_s^2 \, S^\dagger S - \lambda_s \left(S^\dagger S  \right)^2 \,.
\end{align}

Once the Higgs singlet aquires a vacuum expectation value (vev), it will spontaneously break the $U(1)'$ symmetry, thus giving mass to the $Z'$ gauge boson and the DM particle. After symmetry breaking, we obtain the following Lagrangian (defining $S = 1/\sqrt{2} (s + w)$ and using $g_S \equiv g'q_S = -2g_\text{DM}^A$)
\begin{align}
 \mathcal{L} = & \frac{i}{2} \bar{\psi} \slashed{\partial} \psi - \frac{1}{4} F'^{\mu\nu}F'_{\mu\nu} - \frac{1}{2} g_\text{DM}^A Z'^\mu \bar{\psi} \gamma^5 \gamma_\mu \psi - \frac{m_\text{DM}}{2} \bar{\psi} \psi - \frac{y_\text{DM}}{2\sqrt{2}} s \bar{\psi} \psi \nonumber \\
 & + \frac{1}{2} m_{Z'}^2 \, Z'^\mu Z'_\mu + \frac{1}{2} \partial^\mu s \partial_\mu s + 2 (g_\text{DM}^A)^2 \, Z'^\mu Z'_\mu (s^2 + 2\,s\,w) 
+ \frac{\mu_s^2}{2} (s+w)^2 - \frac{\lambda_s}{4} (s+w)^4 \; ,
\end{align}
with $F'^{\mu\nu} = \partial^\mu Z'^\nu - \partial^\nu Z'^\mu$ and
\begin{equation} \label{eq:masses}
  m_\text{DM} = \frac{1}{\sqrt{2}} \, y_\text{DM} \, w\,,\quad
  m_{Z'} \approx 2 g_\text{DM}^A \, w \,.  
\end{equation}
If the SM Higgs is charged under the $U(1)'$ the $Z'$ mass will receive an
additional contribution from the SM Higgs vev, see eq.~\eqref{eq:mZpr}
below. Electroweak precisison data requires that this
  contribution is small, and therefore we neglect this term in
  eq.~\eqref{eq:masses} and for the rest of this subsection.
Note that without loss of
generality we can choose $w$ and $y_\text{DM}$ to be real (ensuring
real masses) by absorbing complex phases in the field definitions for
$S$ and $\psi$.\footnote{This will no longer be true if we allow for
  an explicit mass term for $\psi$. In this case the relative phase
  between $y_\text{DM}$ and the mass term is physical (see
  e.g.~\cite{LopezHonorez:2012kv}). Here we do not allow for an
  explicit mass term and we assume that the vev of the singlet is the
  only source of $U(1)'$ symmetry breaking.}

As discussed above, the mass of the additional Higgs particle must satisfy
\begin{equation}
m_s < \frac{\pi \, m_{Z'}^2}{(g^A_\text{DM})^2 \, m_\text{DM}}
\end{equation}
in order for perturbative unitarity to be satisfied, which when substituting the masses of the $Z'$ and DM becomes
\begin{equation}
m_s < \frac{4\sqrt{2} \pi w}{y_\text{DM}} \;.
\end{equation}
Once we include such a new particle coupling to the $Z'$, however, there are additional scattering processes such as $s s \rightarrow s s$ that need to be taken into account when checking perturbative unitarity~\cite{Basso:2011na}. 
Here we consider the scattering of the states $ss/\sqrt{2}$ and $Z'_L Z'_L/\sqrt{2}$.
In the limit $\sqrt{s} \gg m_s \gg m_{Z'}$, the $J=0$ partial wave of the scattering matrix takes the form~\cite{Lee:1977eg} \begin{equation}
  \lim_{\sqrt{s} \rightarrow \infty}  \mathcal{M}^0_{if}
  = - \frac{(g^A_\text{DM})^2 m_s^2}{8 \pi m_{Z'}^2} 
\begin{pmatrix}
3 & 1\\
1 & 3
\end{pmatrix} \; .
\end{equation}
Partial wave unitarity requires the real part of the largest eigenvalue, which corresponds to the eigenvector $(ss + Z'_L Z'_L)/2$, to be smaller than $1/2$. We hence obtain the inequality
\begin{equation}
 m_s \leq \frac{\sqrt{\pi} \, m_{Z'}}{g_\text{DM}^A} = \sqrt{4 \pi} w \; .
 \label{eq:perturb}
\end{equation}
This inequality together with eq.~(\ref{eq:DMmass}) gives a stronger bound on the Higgs mass than the one obtained in eq.~(\ref{eq:higgsmass}). In other words, the bound in~(\ref{eq:higgsmass}) can never actually be saturated in this UV completion. We note that eqs.~\eqref{eq:DMmass} and \eqref{eq:perturb} can be unified to
\begin{equation}\label{eq:bound_w}
  \sqrt{\pi} \, \frac{m_{Z'}}{g_\text{DM}^A} \ge \text{max}\left[ m_s ,
    \sqrt{2} m_\text{DM}\right] \,.
\end{equation}

\subsection{Implications for the visible sector}

For the discussion above we only needed to consider the DM part of the Lagrangian. Let us now also look at the coupling to the SM, see e.g.~\cite{Carena:2004xs}. The interactions between SM states and the new $Z'$ gauge boson can be written as
\begin{align}
  \mathcal{L}'_\text{SM} = &
  \left[ (D^\mu H)^\dagger (-i \, g' \, q_H \, Z'_\mu \, H) + \text{h.c.} \right] +
  g'^2 \, q_H^2 \, Z'^\mu Z'_\mu \, H^\dagger H \nonumber \\
  & - \sum_{f = q,\ell,\nu} g' \, Z'^\mu \, \left[ q_{f_L} \, \bar{f}_L \gamma_\mu f_L  + q_{f_R} \, \bar{f}_R \gamma_\mu f_R \right] \; ,
\end{align}
where $D^\mu$ denotes the SM covariant derivative.  We can now
immediately write down a list of relations between the different
charges $q$ required by gauge invariance of the SM Yukawa
terms:\footnote{If right-handed neutrinos exist their charge
  $q_{\nu_R}$ would be constrained by $q_H = q_{\ell_L} -
  q_{\nu_R}$ to allow for a Yukawa term with the lepton doublet. In the following
  we assume that if right-handed neutrinos exist they are heavy enough
  to decouple from all relevant phenomenology.}
\begin{align} \label{eq:charges}
q_H = q_{q_L} - q_{u_R} = q_{d_R} - q_{q_L} = q_{e_R} - q_{\ell_L} \; .
\end{align}

After electroweak symmetry breaking, we obtain
\begin{align}
  \mathcal{L}'_\text{SM} & =
  \frac{1}{2} \frac{e \, g' \, q_{H}}{ s_\mathrm{W} \, c_\mathrm{W}} (h+v)^2 \, Z^\mu Z'_\mu
  + \frac{1}{2} g'^2 \, q_{H}^2 \, (h+v)^2 \, Z'^\mu Z'_\mu \nonumber \\
  & \quad -
  \sum_{f = q,l,\nu} \frac{1}{2} g' Z'^\mu \, \bar{f}
  \left[ (q_{f_R} + q_{f_L}) \gamma_\mu + (q_{f_R} - q_{f_L}) \gamma_\mu \gamma^5 \right] f \; .
  \label{eq:LprSM}
\end{align}
Comparing the second line of eq.~\eqref{eq:LprSM} with
eq.~\eqref{eq:L_VA} we can read off the vector and axial vector
couplings of the fermions:
\begin{equation}\label{eq:g_f_VA}
  g_f^V = \frac{1}{2}g'(q_{f_R} + q_{f_L}) \,,\quad
  g_f^A = \frac{1}{2}g'(q_{f_R} - q_{f_L}) \,.
\end{equation}

It is well known that a $U(1)'$ under which only SM fields are charged is in general anomalous,
unless the SM fields have very specific charges (e.g.\ $U(1)_{B-L}$ is anomaly free). The relevant anomaly coefficients can e.g.\ be found in \cite{Carena:2004xs}. The presence of these anomalies implies that the theory has to include new fermions to cancel the anomalies. While these fermions can be vectorlike with respect to the SM, they will then need to be chiral with respect to the $U(1)'$. The mass of the additional fermions is therefore constrained by the breaking scale of the $U(1)'$. In particular, the bound from eq.~\eqref{eq:DMmass} applies to these fermions as well and therefore they cannot be decoupled from the low-energy theory.

It is however interesting to note that the anomaly involving two gluons and a $Z'$ is proportional to
\begin{equation}
A_{ggZ'}  =   3 \left( 2q_{q_L} - q_{u_R} - q_{d_R} \right) \; ,
\end{equation}
which always vanishes if we restrict the charges based on gauge invariance of the Yukawa couplings (see eq.~\eqref{eq:charges}). This implies that no new coloured states are needed to cancel the anomalies, greatly reducing the 
sensitivity of colliders to these new states.\footnote{This conclusion is in disagreement with the observations made in~\cite{Hooper:2014fda}.} In any case, there are many different possibilities for cancelling the anomalies via new fermions. While the existence of additional fermions will lead to new signatures, a detailed investigation of these is beyond the scope of this work.

If the SM Higgs is charged under $U(1)'$ ($q_H \neq 0$) the mass of
the $Z'$ receives a contribution from both Higgses:
\begin{equation}\label{eq:mZpr}
  m_{Z'}^2 = (g'q_H v)^2 + 4(g_\text{DM}^A w)^2  \,,
\end{equation}
and we obtain a mass mixing term of the form $\delta m^2 \, Z^\mu Z'_\mu$ with
\begin{align}
\delta m^2 =  \frac{1}{2} \frac{e \, g' \, q_{H}}{ s_\mathrm{W} \, c_\mathrm{W}} v^2 \; ,
\end{align}
where $s_\mathrm{W} \,(c_\mathrm{W})$ is the sine (cosine) of the Weinberg angle.

As we are going to discuss below, electroweak precision data requires
$|\delta m^2| \ll |m_Z-m_{Z'}|$ (see also
App.~\ref{app:Z}). Using $m_Z = e v / (2 s_\mathrm{W} c_\mathrm{W})$, $g'q_s =
-2g_\text{DM}^A$, and neglecting order one factors this requirement
implies either $g'q_H \ll e$ or $q_s w \gg v$. In the parameter regions of interest it follows from those conditions
  that the first term in eq.~\eqref{eq:mZpr} is small and hence the mass of
  the $Z'$ is dominated by the vev of the dark Higgs. Taking into account
eqs.~\eqref{eq:charges} and \eqref{eq:g_f_VA}, the condition $|\delta m^2| \ll |m_Z - m_{Z'}|$ then implies either small axial couplings ($g_f^A \ll 1$) or $m_{Z'} \gg m_Z$. We are going to present more quantitative results in the next section and discuss a number of interesting experimental signatures resulting from the new interactions due to eq.~\eqref{eq:LprSM}.

To conclude this section, it should be noted, that  the Lagrangian introduced above is UV-complete (up to anomalies) and  gauge invariant but does not correspond to the most general realization of this model. In particular the term 
\begin{equation}
  \mathcal{L} \supset - \lambda_{hs} (S^*S)(H^\dagger H) \,,
\end{equation}
which will lead to mixing between the SM Higgs $h$ and the dark Higgs $s$ 
can be expected to be present at tree level. Furthermore, the term
\begin{equation}\label{eq:kin-mix}
\mathcal{L} \supset -\frac{1}{2} \sin \epsilon F'^{\mu\nu} B_{\mu\nu} \; ,
\end{equation}
which generates kinetic mixing between the $Z'$ and the $Z$-boson, respects all symmetries of the Lagrangian. It can be argued that $\epsilon$
might vanish at high scales in certain UV-completions, but even in this case
kinetic mixing is necessarily generated at the one-loop level and can have a substantial impact on EWPT. We will return to these issues and the resulting phenomenology of the model in sections~\ref{sec:vector} and \ref{sec:higgsmixing}. For the moment, however, we are going to neglect these additional effects and focus on the impact of $\delta m \neq 0$, which necessarily leads to mass mixing between the neutral gauge bosons in the case of non-vanishing axial couplings.

\section{Non-zero axial couplings to SM fermions}
\label{sec:axial}

Let us start with the case that axial couplings on the SM side are
non-vanishing. An immediate consequence is that the SM Higgs is
charged under the $U(1)'$, which follows from eqs.~\eqref{eq:charges}
and \eqref{eq:g_f_VA} for $g_f^A \neq 0$. Note that these equations also imply that it is inconsistent to set the vectorial couplings for all quarks equal to zero. For example, if we impose that the vectorial couplings of up quarks vanish, i.e. $g_u^V = 0$, eq.~\eqref{eq:g_f_VA} implies $q_{u_R} = - q_{q_L}$, which using eq.~\eqref{eq:charges} leads to $g_d^V = 2 g' q_{q_L}$. In the following, whenever $g^A_f \neq 0$, we always fix $g^V_f = g^A_f$, which corresponds to setting $q_{q_L} = q_{\ell_L} = 0$.

Furthermore, eq.~\eqref{eq:charges} requires that $Z'$ couplings are
flavour universal and leptons couple with the same strength to the
$Z'$ as quarks. This conclusion could potentially be modified by considering an extended Higgs sector, e.g.\ a two-Higgs-doublet  model. Here we focus on the simplest case where a single Higgs doublet generates all SM fermion masses. This implies that the leading search channel at the
LHC will be dilepton resonances, which give severe constraints.  In principle
also electron-positron colliders can constrain this scenario
efficiently.  Limits on a $Z'$ lighter than 209~GeV derived from LEP
data imply $g \lesssim 10^{-2}$ \cite{Agashe:2014kda} (see also
\cite{Appelquist:2002mw,LEP:2003aa}). We do not include LEP constraints here since other constraints will turn out to be at least equally strong.

For general couplings, the partial decay width of the mediator into SM fermions is given by
\begin{equation}
\Gamma(Z'\rightarrow f\bar{f}) = \frac{m_{Z'} N_c}{12\pi} 
 \sqrt{1-\frac{4 m_f^2}{m_{Z'}^2}} \, \left[(g^{V}_{f})^2+(g^{A}_{f})^2 + \frac{m_f^2}{m_{Z'}^2}\left(2 (g^{V}_{f})^2 - 4 (g^{A}_{f})^2\right)  \right]  \; ,
\end{equation}
where $N_c = 3 \;(1)$ for quarks (leptons). The decay width into DM pairs is
\begin{equation}
\Gamma(Z'\rightarrow \psi\psi) = \frac{m_{Z'}}{24\pi} 
 (g^{A}_\text{DM})^2 \left(1-\frac{4 m_\text{DM}^2}{m_{Z'}^2}\right)^{3/2}  \; .
\end{equation}
Consequently, for $m_\text{DM} \ll m_{Z'}$ and $g^A_{\ell} = g^A_{q} \ll g^A_\text{DM}$ the branching ratio into $\ell = e,\,\mu$ is given by $\text{BR}(R\rightarrow \ell \ell) \approx 8 (g^A_{\ell})^2 / (g^A_\text{DM})^2$. For $m_\text{DM} > m_{Z'} / 2$, on the other hand, the branching ratio is given by $\text{BR}(R\rightarrow \ell \ell) \approx 0.08\text{--}0.10$ depending on the ratio $m_{Z'} / m_t$.

\begin{figure}[tb]
\centering
\includegraphics[height=0.3\textheight]{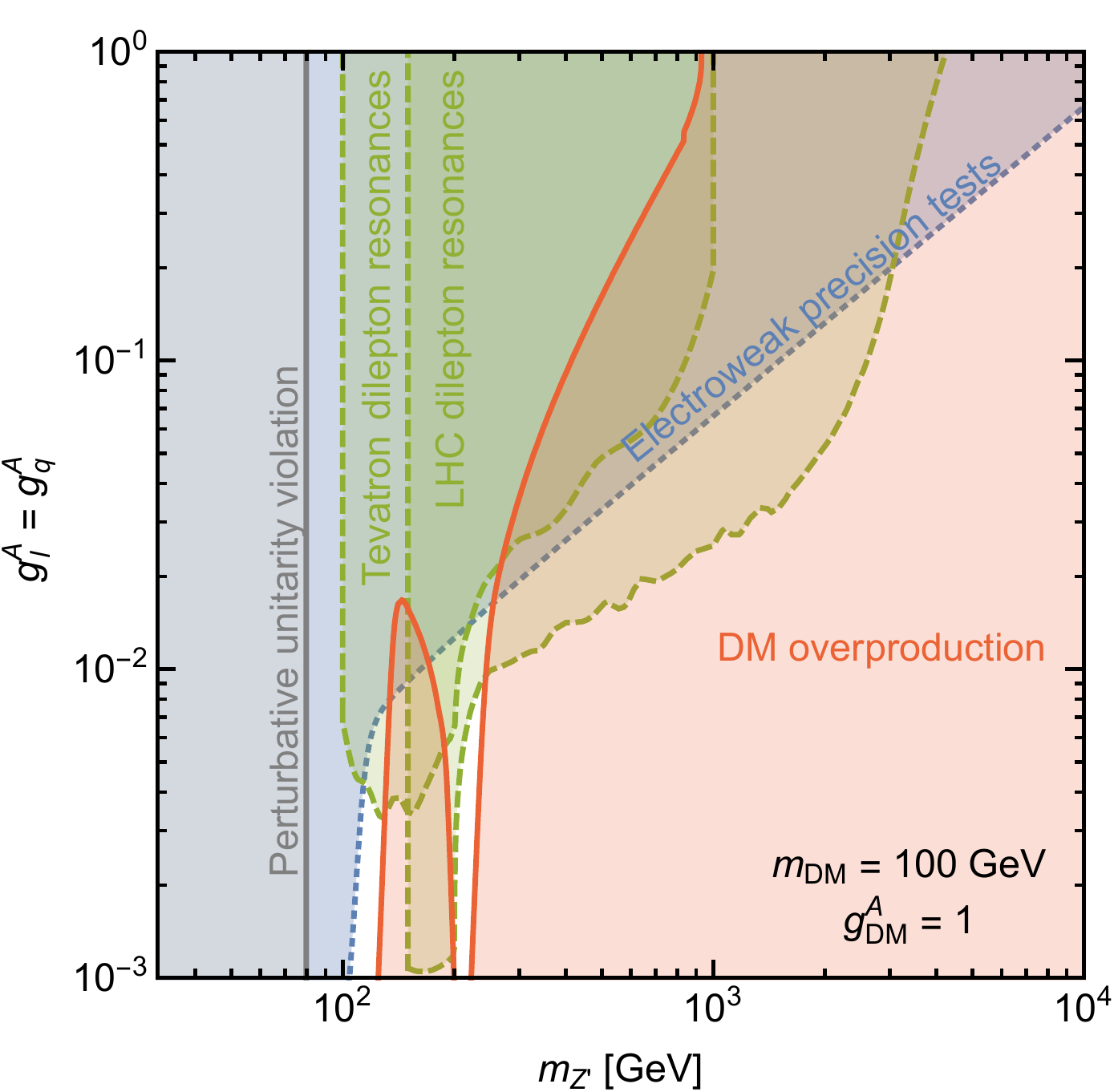}\qquad
\includegraphics[height=0.3\textheight]{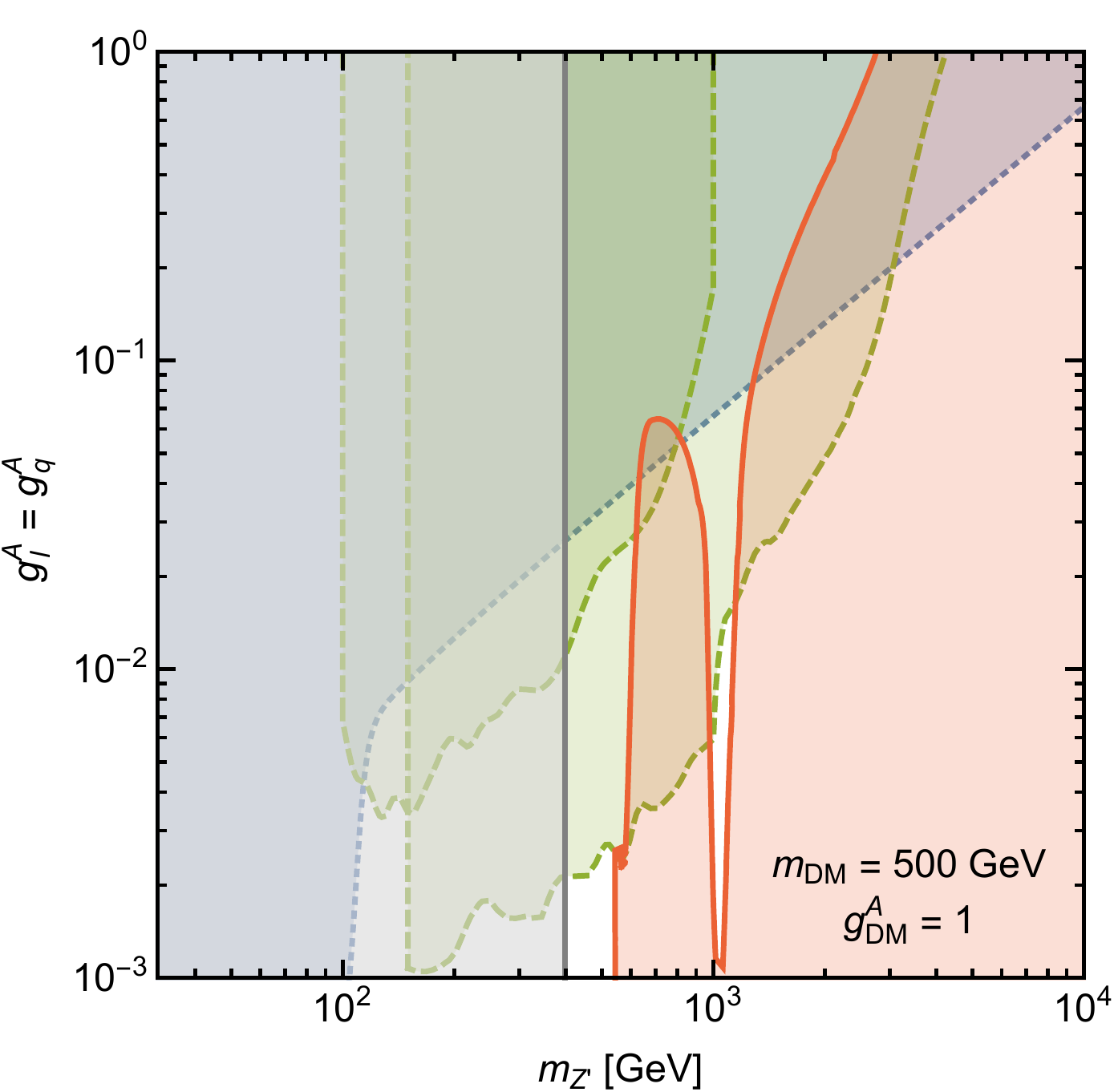}
\caption{Axial+Vector(SM)--Axial(DM): 
  Parameter space forbidden by constraints from dilepton resonance searches from ATLAS (light green, dashed) and Tevatron (dark green, dashed), electroweak precision observables (blue, dotted) and DM overproduction (red, solid) in the $m_{Z'}-g_{q,l}^A$ parameter plane for two exemplary DM masses 100 GeV (left) and 500~GeV (right). In the shaded region to the left of the vertical grey line the $Z'$-mass violates the bound from perturbative unitarity from eq.~\eqref{eq:DMmass}.}
\label{fig:axialaxial}
\end{figure}

We implement the latest ATLAS dilepton search \cite{Aad:2014cka},
complemented by a Tevatron dilepton search \cite{Jaffre:2009dg}
  for the low mass region, and show the resulting bounds in
figure~\ref{fig:axialaxial}. One can see that the bounds strongly
depend on the assumed branching ratio of the $Z'$. As a conservative
limiting case we show $g_\text{DM}^A=1$ and $m_\text{DM}=100$ GeV,
which leads to a rather large branching fraction into DM and hence
suppressed bounds. The second benchmark, $m_\text{DM}=500$~GeV,
allows for $Z'$ decays to DM only for rather heavy $Z'$s, leading to correspondingly
more restrictive dilepton constraints. Overall the bounds turn out to be very stringent
and the $Z'$ coupling to leptons and quarks needs to be significantly
smaller than unity for $100 \; \text{GeV} \lesssim m_{Z'} \lesssim 4 \;
\text{TeV}$, so that dijet constraints are basically irrelevant in
this case given that $g_q=g_l$.

The fact that the SM Higgs is charged also implies potentially large corrections to electroweak precision observables.
In particular we obtain the non-diagonal mass term $\delta m^2 \,
Z^\mu Z'_\mu$ leading to mass mixing between the SM $Z$ and the new
$Z'$.  The diagonalisation required to obtain mass eigenstates is discussed in the appendix. 
In the absence of kinetic mixing between the $U(1)'$ and the SM $U(1)$ gauge bosons ($\epsilon = 0$), the resulting effects can be expressed in
terms of the mixing parameter $\xi = \delta m^2 /(m_Z^2-m_{Z'}^2)$ (see eq.~\eqref{eq:xi_def} in the appendix with $\epsilon=0$). In particular, we can calculate the constraints from electroweak
precision measurements, which are encoded in the $S$ and $T$
parameters. To quadratic order in $\xi$ we find~\cite{Frandsen:2011cg}
\begin{align}
  \alpha S = & - 4 c_\mathrm{W}^2 s_\mathrm{W}^2 \xi^2 
  \; , \nonumber\\
  \alpha T = & \xi^2\left(\frac{m_{Z'}^2}{m_{Z}^2}-2\right) \; , \label{eq:ST}
\end{align}
where $\alpha=e^2/4\pi$. The resulting bounds are shown in figure~\ref{fig:axialaxial}.  To infer
our bounds we use the $90\%$ CL limit on the $S$ and $T$ parameters as
given in \cite{Agashe:2014kda}.

Note that the bound from electroweak precision data is
  completely independent of the $Z'$ couplings to the DM as well as
  the DM mass. Hence, the same bound would apply also in the
  case of Dirac DM with vector couplings to the $Z'$. Note however
  that since vectorial couplings to quarks are necessarily non-zero if $g_q^A \neq 0$ (see
    eqs.~\eqref{eq:charges}, \eqref{eq:g_f_VA}), there will be very
  stringent bounds from direct detection experiments on any model with
  $g^V_\text{DM} \neq 0$ due to unsuppressed spin-independent
    scattering. For Majorana DM, the vectorial coupling always
  vanishes and the constraints from direct detection are much weaker.

\begin{figure}[t]
\centering
\includegraphics[height=0.3\textheight]{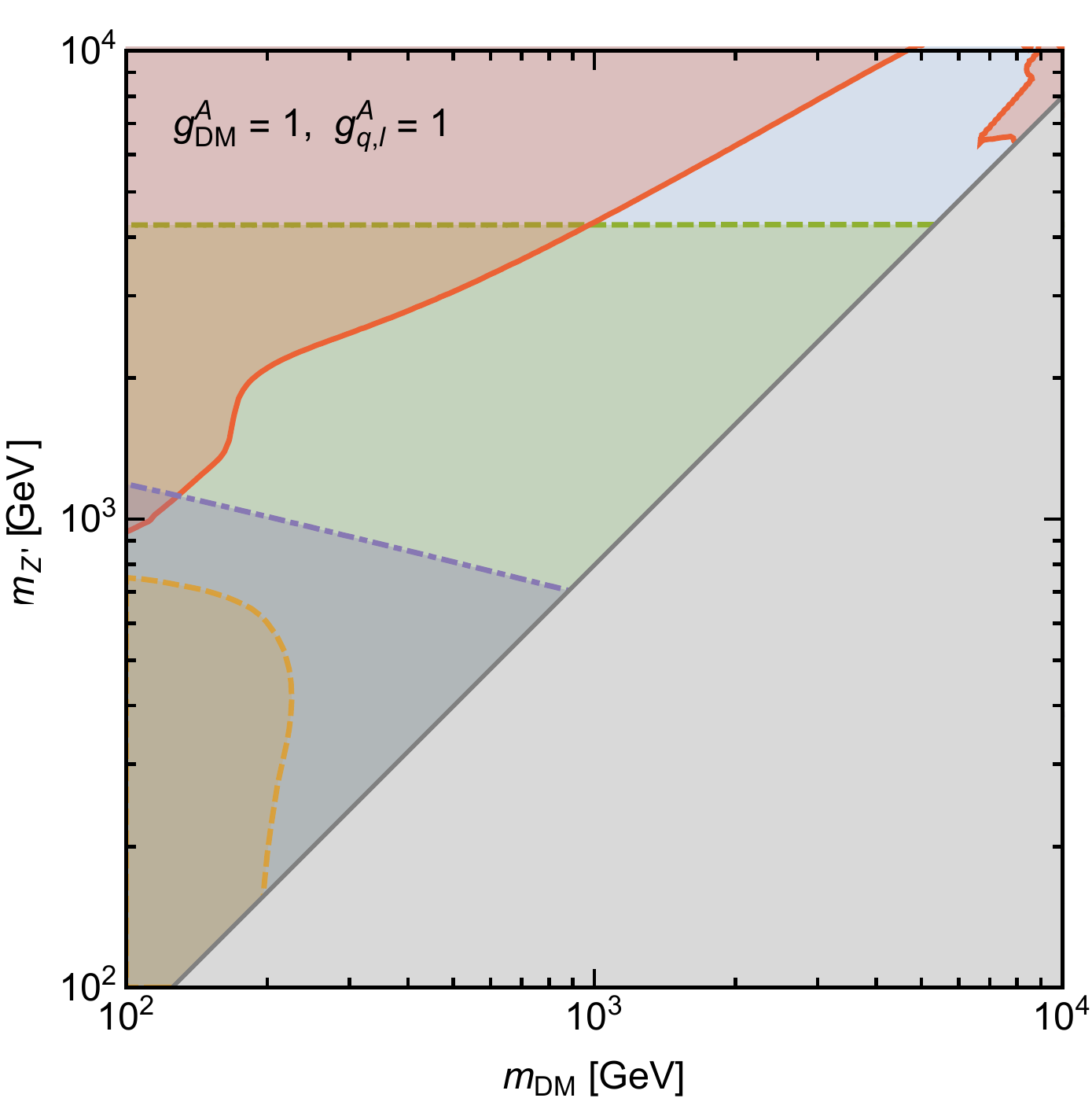}\qquad
\includegraphics[height=0.3\textheight]{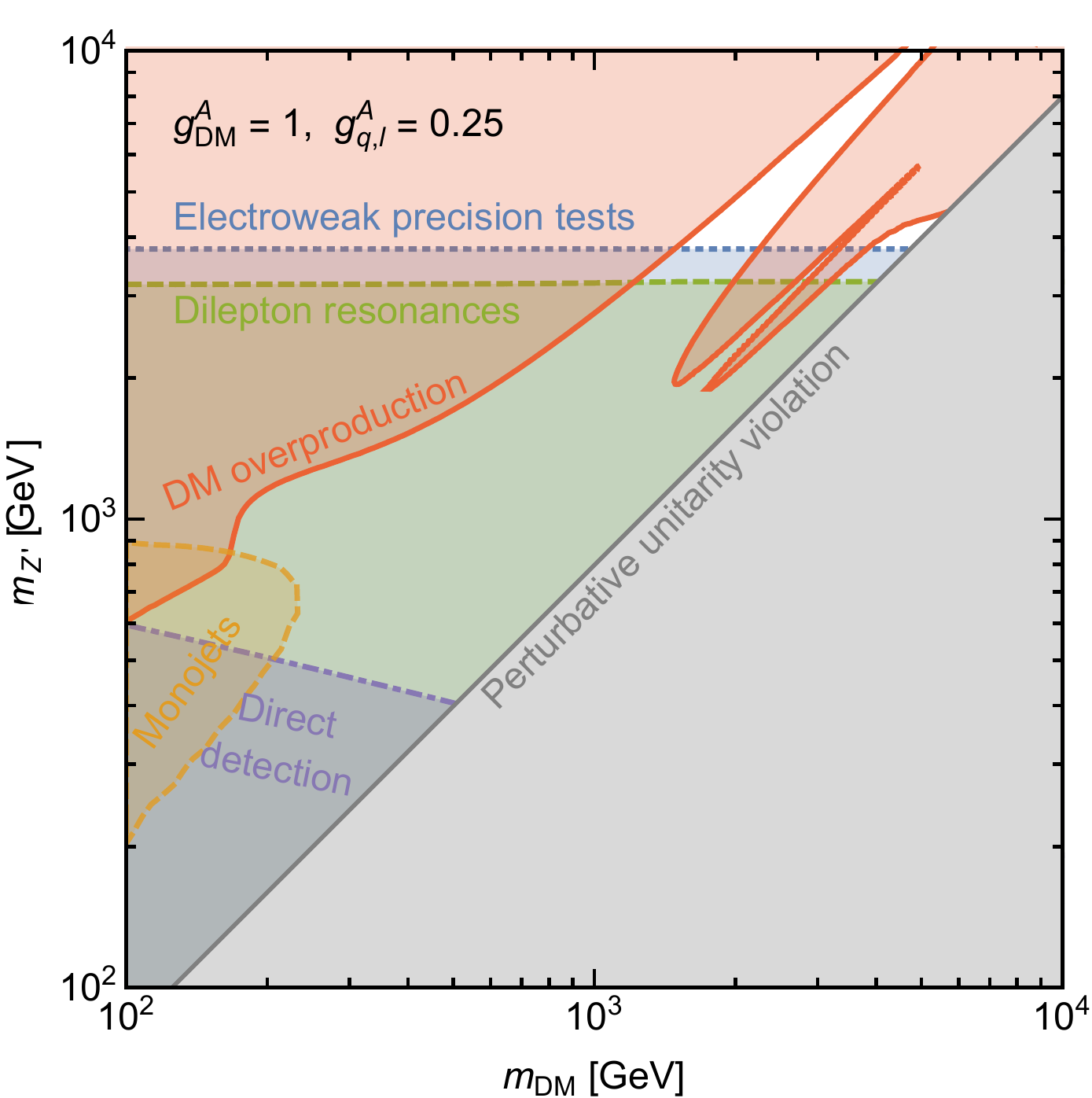}
\includegraphics[height=0.3\textheight]{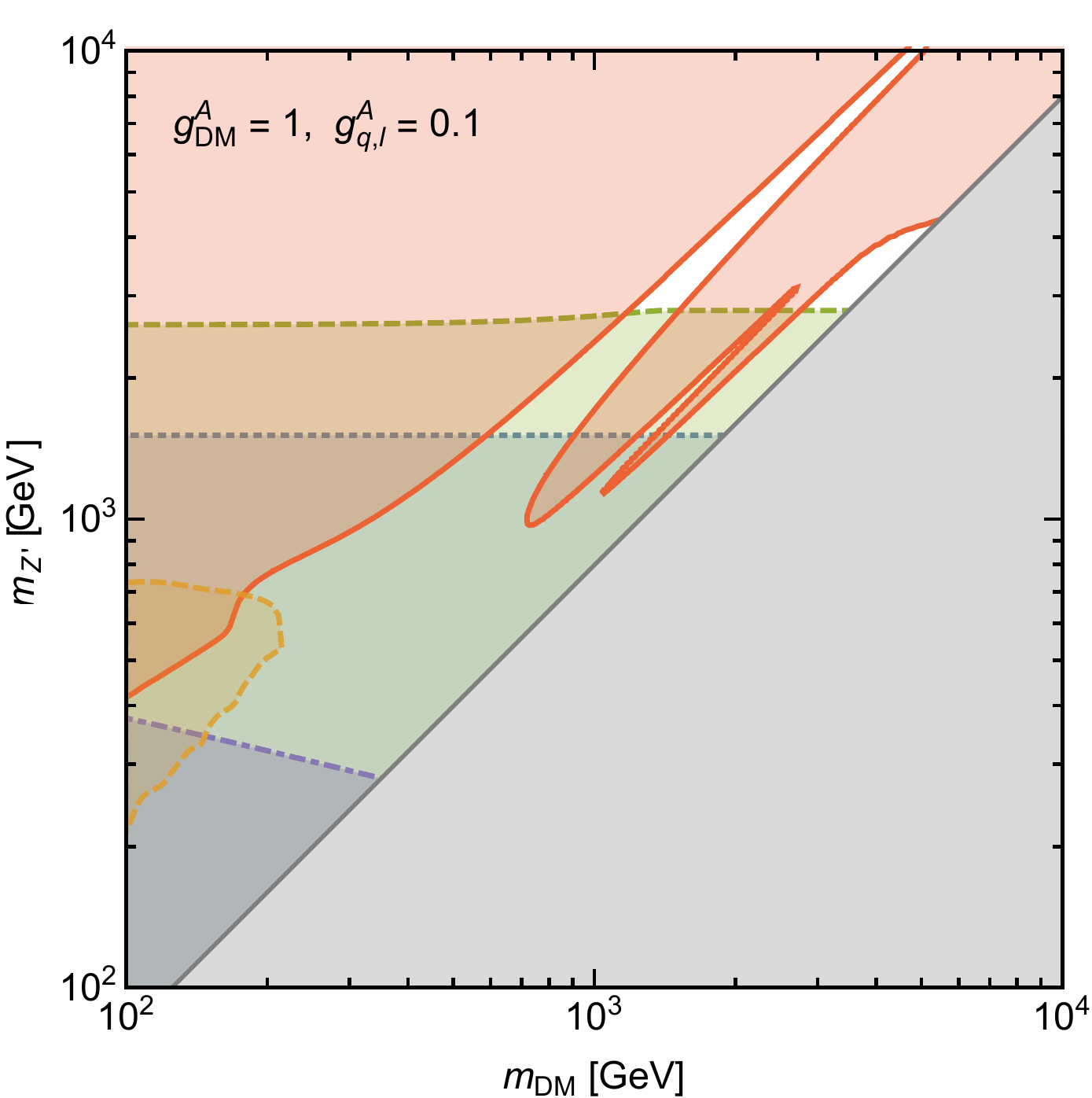}\qquad
\includegraphics[height=0.3\textheight]{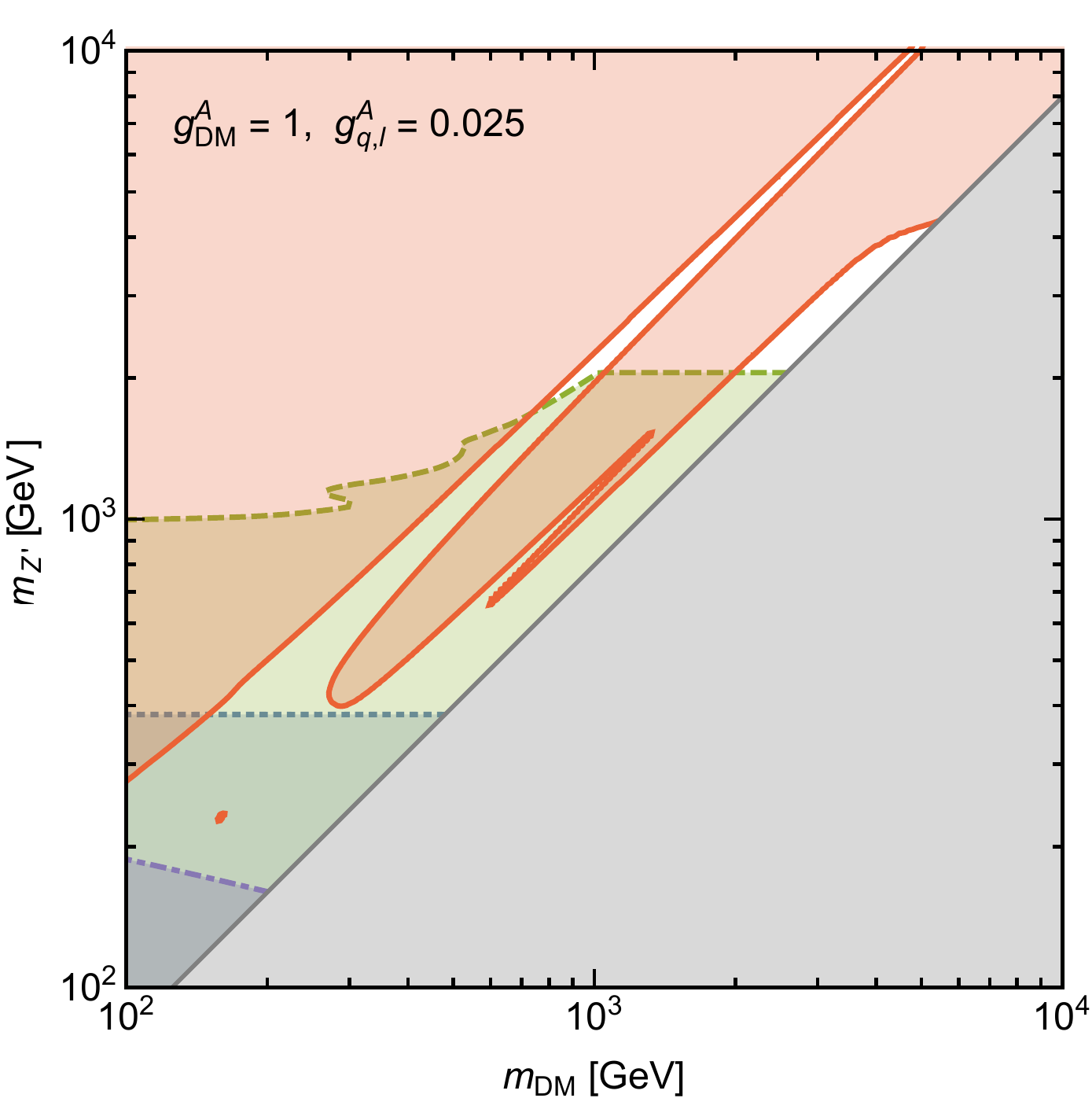}
\caption{Axial+Vector(SM)--Axial(DM): Parameter space forbidden by constraints from dilepton resonance searches (green, dashed) and electroweak precision observables (blue, dotted) in the $m_\text{DM}-m_{Z'}$ plane for four different sets of couplings. We also show the regions excluded by DM overproduction (red), direct detection bounds (purple, dot-dashed) and the parameter space where perturbative unitarity is violated (grey). For the relic density calculation we have assumed that the mass of the hidden sector Higgs saturates the unitarity bound.}
\label{fig:axialaxial2}
\end{figure}

In figure~\ref{fig:axialaxial2} we show the constraints from
electroweak precision data as well as LHC dilepton searches in the
$m_\text{DM}-m_{Z'}$ plane for different values of the axial vector
coupling to fermions. In the lower right corner of the plots (grey area) the
perturbative unitarity condition from eq.~\eqref{eq:DMmass} is
violated. We also show the region excluded by direct detection
searches (dark region in the lower left corners). For the axial-axial
couplings DM-nucleus scattering proceeds through spin-dependent
interactions, with a scattering cross section given by
\begin{equation}
\sigma^\text{SD}_N = \frac{3 \, a_N^2 \, (g^A_\text{DM})^2 \, (g^A_q)^2}{\pi} \frac{\mu^2}{m_{Z'}^4} \; ,
\end{equation}
where $\mu$ is the DM-nucleon reduced mass, $N = p,\,n$ and $a_p = -a_n = 1.18$ is the effective nucleon coupling~\cite{Agashe:2014kda}. 
This is the dominant contribution in this case as the vector-axial coupling combination is even further suppressed.
In the plots we show the bound on the spin-dependent scattering cross section that can be calculated from the published LUX results~\cite{Akerib:2013tjd}, following the method described in~\cite{Feldstein:2014ufa}. We observe
that in this case direct detection is never competitive with other
constraints.

The red solid curves in Figs.~\ref{fig:axialaxial} and
  \ref{fig:axialaxial2} show the parameter values that lead to the correct relic abundance.  In
order to calculate the relic abundance we have implemented the model
in micrOMEGAs\_v4~\cite{Belanger:2014vza}, assuming that the mass of the Higgs
singlet saturates the unitarity bound and setting the mixing with the
SM Higgs to zero.\footnote{Note that since $\xi$ can be large in some regions of parameter space, it is not a good approximation to expand the annihilation cross section in $\xi$. We therefore use the exact expression for the mixing between the neutral gauge bosons in terms of $\epsilon$ and $\delta m^2$ as derived in the appendix.}
In the regions shaded in red (to the right/above the solid curve) there is overproduction of DM. In
this region additional annihilation channels are required to avoid
overclosure of the Universe, since the interactions provided by the
$Z'$ are insufficient to keep DM in thermal equilibrium long
enough. Such additional interactions could be obtained for instance from
the scalar mixing discussed in section~\ref{sec:higgsmixing}. Conversely, to the left/below the red
solid curve the model does not provide all of the DM matter in the
Universe, since the annihilation rate is too high. 

Let us briefly discuss the various features that can be observed in the relic abundance curve. First there is a significant decrease of the predicted abundance as the DM mass crosses the top-quark threshold, $m_\chi > m_t$, resulting from the fact that the $s$-wave contribution to the annihilation cross section is helicity suppressed and hence annihilation into top-quarks becomes the dominant annihilation channel as soon as it is kinematically allowed. The second feature occurs at $m_\chi \sim m_{Z'}$ and reflects the resonant enhancement of the annihilation process $\chi \chi \rightarrow q \bar{q}$ as the mediator can be produced on-shell. A third visible feature is a very narrow resonance at $2 \, m_\chi \sim m_s = \sqrt{\pi} \,  m_{Z'} / g^A_\text{DM}$ due to a resonant enhancement of the process $\chi \chi \rightarrow s \rightarrow Z' Z$. The position and magnitude of this effect depends on the mass of the dark Higgs, which has been (arbitrarily) fixed to saturate the unitarity bound. However, even for this extreme choice, it turns out to give a non-negligible contribution to the relic abundance. For $m_\chi > m_{Z'}$ direct annihilation into two mediators becomes possible, leading to a significant decrease of the predicted relic abundance. Finally, the fact that the relic abundance curve in figure~\ref{fig:axialaxial2} touches the unitarity bound for high DM masses reflects the well-known unitarity bound on the mass of a thermally produced DM particle \cite{Griest:1989wd}.

All in all we find the case with non-vanishing axial couplings on the
SM side to be strongly constrained by dilepton searches as well as
electroweak precision observables, implying that in a UV complete
model this is where a signal should first be seen. For comparison, we show recent bounds from LUX as well as from the CMS monojet search~\cite{Khachatryan:2014rra}.\footnote{To interpret the CMS results in the context of our model, we implement our model in \texttt{Feynrules~v2}~\cite{Alloul:2013bka} and simulate the monojet signal with \texttt{MadGraph~v5}~\cite{Alwall:2014hca} and \texttt{Pythia~v6}~\cite{Sjostrand:2006za}. Imposing a cut on the missing transverse energy of $\slashed{E}_T > 450\:\text{GeV}$, we exclude all parameter points that predict a contribution to the monojet cross section larger than $7.8\:\text{fb}$. We find good agreement between this procedure and an analogous implementation using \texttt{CalcHEP v3}~\cite{Belyaev:2012qa} and \texttt{DELPHES v3}~\cite{deFavereau:2013fsa}.}
We find that these searches, as well as searches for dijet resonances, are not competitive. Note that in figure~\ref{fig:axialaxial2} we assume $g_\text{DM}^A = 1$. We comment on smaller couplings on the DM side later in the context of figure~\ref{fig:smallgA}. Let us now look at the case where axial couplings to
quarks are taken to be zero, which will turn out to be somewhat less constrained.

\section{Purely vectorial couplings to SM fermions}
\label{sec:vector}

Let us now consider the case with purely vectorial couplings on the SM
side, i.e.\ $g^A_q = g^A_\ell = g^A_\nu = 0$. In this case the SM
Higgs does not carry a $U(1)'$ charge and therefore the charges of
quarks and leptons are independent. In particular, it is conceivable
that $g^V_q \gg g^V_\ell$, so that constraints from dilepton resonance
searches can be evaded. Also there can in principle be a flavour dependence
  of the $Z'$ couplings to quarks. Nevertheless, to avoid large flavour-changing neutral currents, we will always assume the
  same coupling for all quark families in what follows~\cite{Abdallah:2015ter}. Finally, in contrast to the case discussed above,
tree-level $Z-Z'$ mass mixing is absent. It therefore seems plausible that
 the $Z'$ is the only state coupling to both the visible
 and the dark sector.
 Nevertheless, as mentioned above, potentially important effects in this scenario can be kinetic mixing of the $U(1)$ gauge bosons as well as effects induced by the dark Higgs, which we are going to discuss below. Let us just mention that all these effects will also be present in the scenario discussed in the previous section.  They are, however, typically less important than the effects of tree-level $Z$-$Z'$ mixing.

We first consider the effects of kinetic mixing between the $Z'$ and the SM hypercharge gauge boson $B$:
\begin{equation}
\mathcal{L} \supset -\frac{1}{2} \sin \epsilon \, F'^{\mu\nu} B_{\mu\nu} \; ,
\end{equation}
where $F'^{\mu\nu} = \partial^\mu Z'^\nu - \partial^\nu Z'^\mu$ and $B^{\mu\nu} = \partial^\mu B^\nu - \partial^\nu B^\mu$. A non-zero value of $\epsilon$ leads to mixing between the $Z'$ and the neutral gauge bosons of the SM (see App.~\ref{app:Z}). As in the case of mass mixing discussed above, there are strong constraints on kinetic mixing from searches for dilepton resonances and electroweak precision observables. 

\begin{figure}[t]
\centering
\includegraphics[height=0.3\textheight]{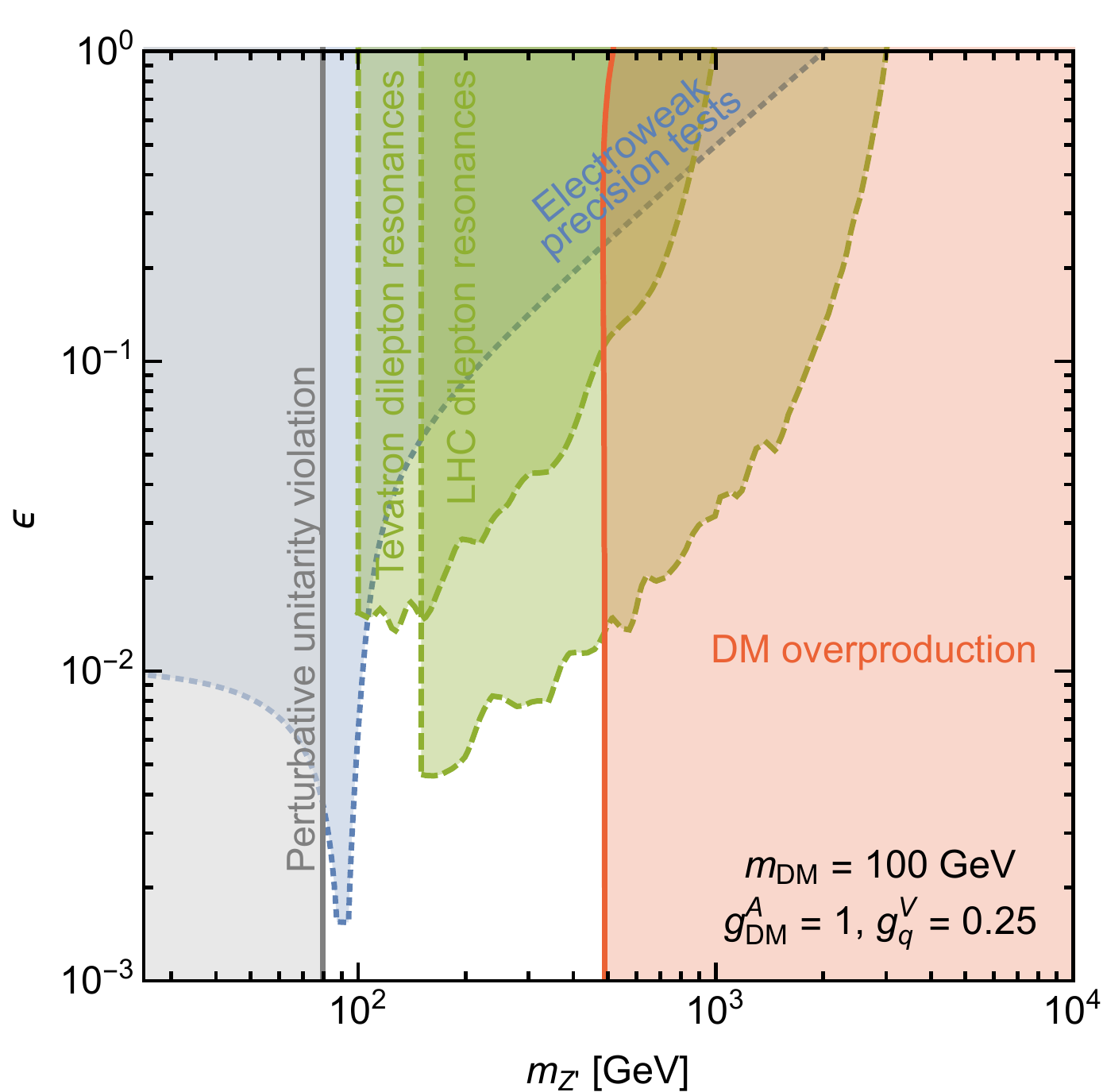}\qquad
\includegraphics[height=0.3\textheight]{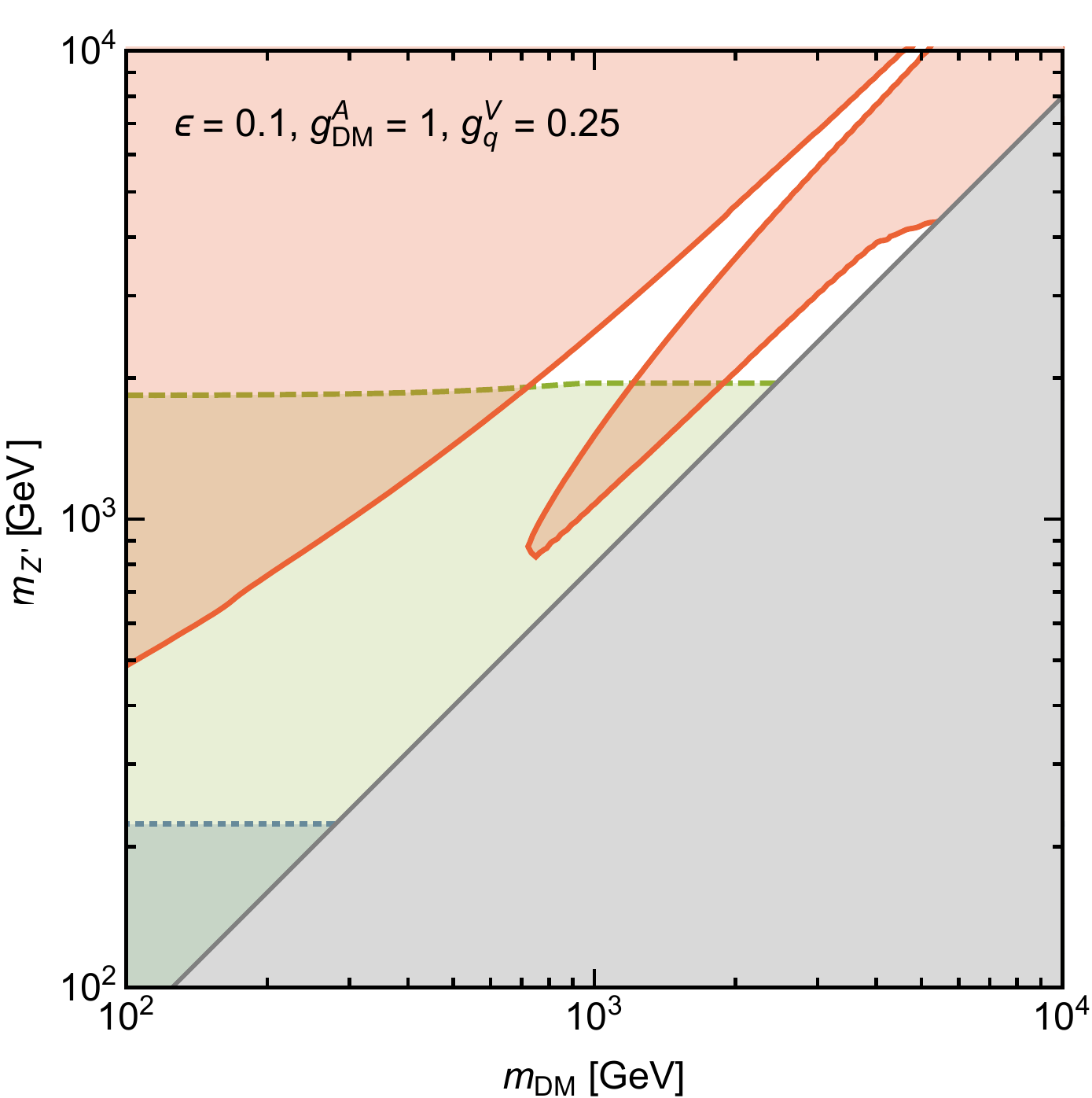}
\caption{Vector(SM)--Axial(DM): Parameter space forbidden by
  constraints from ATLAS and Tevatron dileptons (green, dashed),
  electroweak precision observables (blue, dotted) and relic DM
  overproduction (red, solid) in the $m_{Z'}$-$\epsilon$ parameter
  plane (left) and the $m_\text{DM}$-$m_{Z'}$ parameter plane
  (right). In both panels we
    show the parameter space where perturbative unitarity is violated
    (grey). For the relic density calculation we have assumed that
  the mass of the hidden sector Higgs saturates the unitarity bound.}
\label{fig:va}
\end{figure}

The dilepton couplings induced via the kinetic mixing parameter $\epsilon$ can be inferred from the mixing matrices and are given in the appendix, cf.~eq.~\eqref{eq:dilepton}. The $S$ and $T$ parameters are given by
\begin{align}
  \alpha S = & 4 c_\mathrm{W}^2 s_\mathrm{W} \xi (\epsilon - s_\mathrm{W} \xi) 
  \; , \nonumber\\
  \alpha T = & \xi^2\left(\frac{m_{Z'}^2}{m_{Z}^2}-2\right)+2
  s_\mathrm{W} \xi \epsilon \; , \label{eq:ST2}
\end{align}
where for $\delta m^2 = 0$ the mixing parameter $\xi$ is given by $\xi = m_{Z}^2  s_\mathrm{W} \epsilon / (m_{Z}^2 - m_{Z'}^2)$ at leading order. If $\epsilon$ is sizeable, i.e.\ if mixing is present at tree level, the resulting bounds can be quite strong. This expectation is confirmed in figure~\ref{fig:va}. Note that the relic density curves shown in figure~\ref{fig:va} are basically independent of $\epsilon$, because freeze-out is dominated by direct $Z'$ exchange for the adopted choice of couplings.

\begin{figure}[t]
\centering
\includegraphics[height=0.3\textheight]{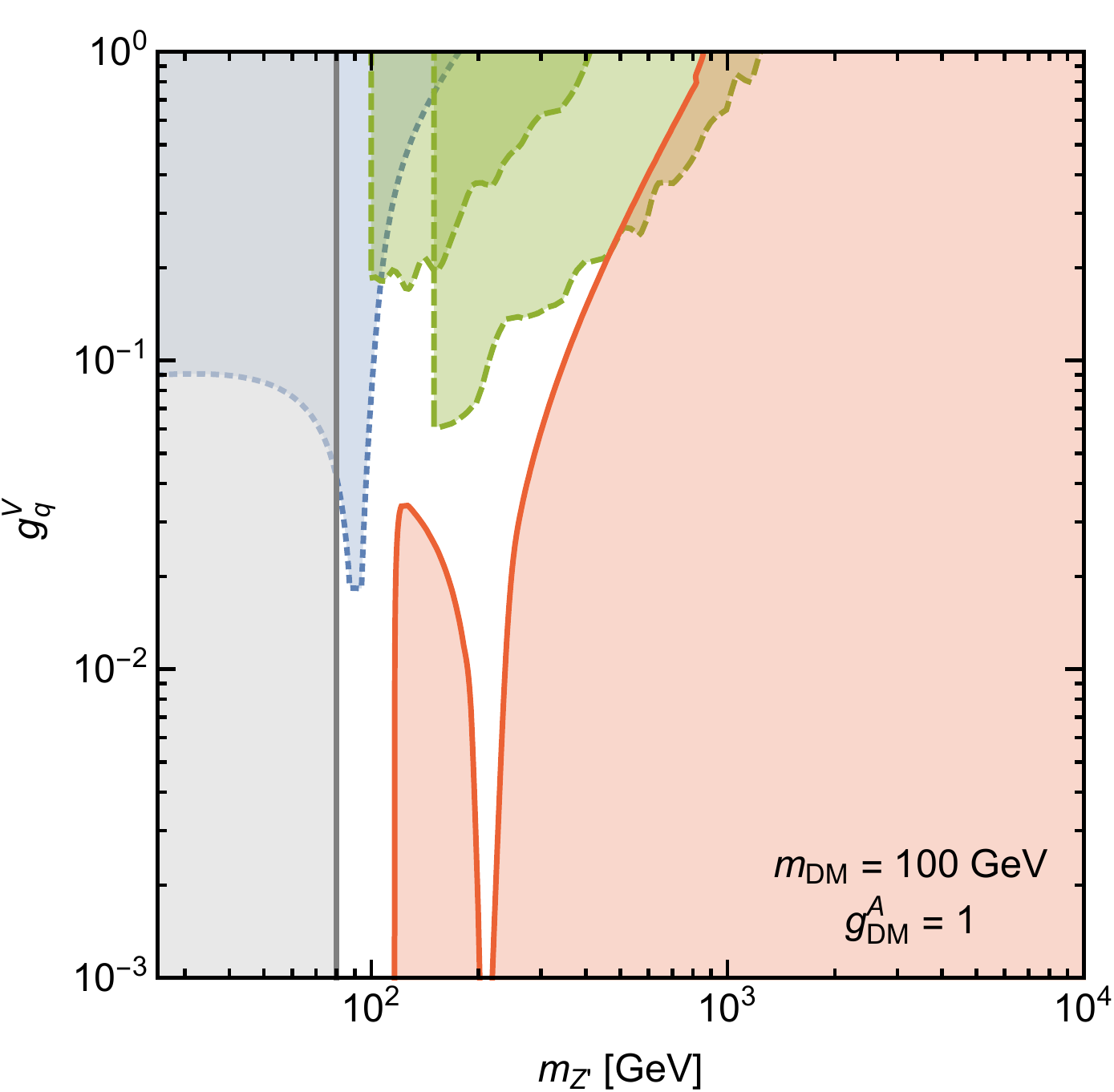}\qquad
\includegraphics[height=0.3\textheight]{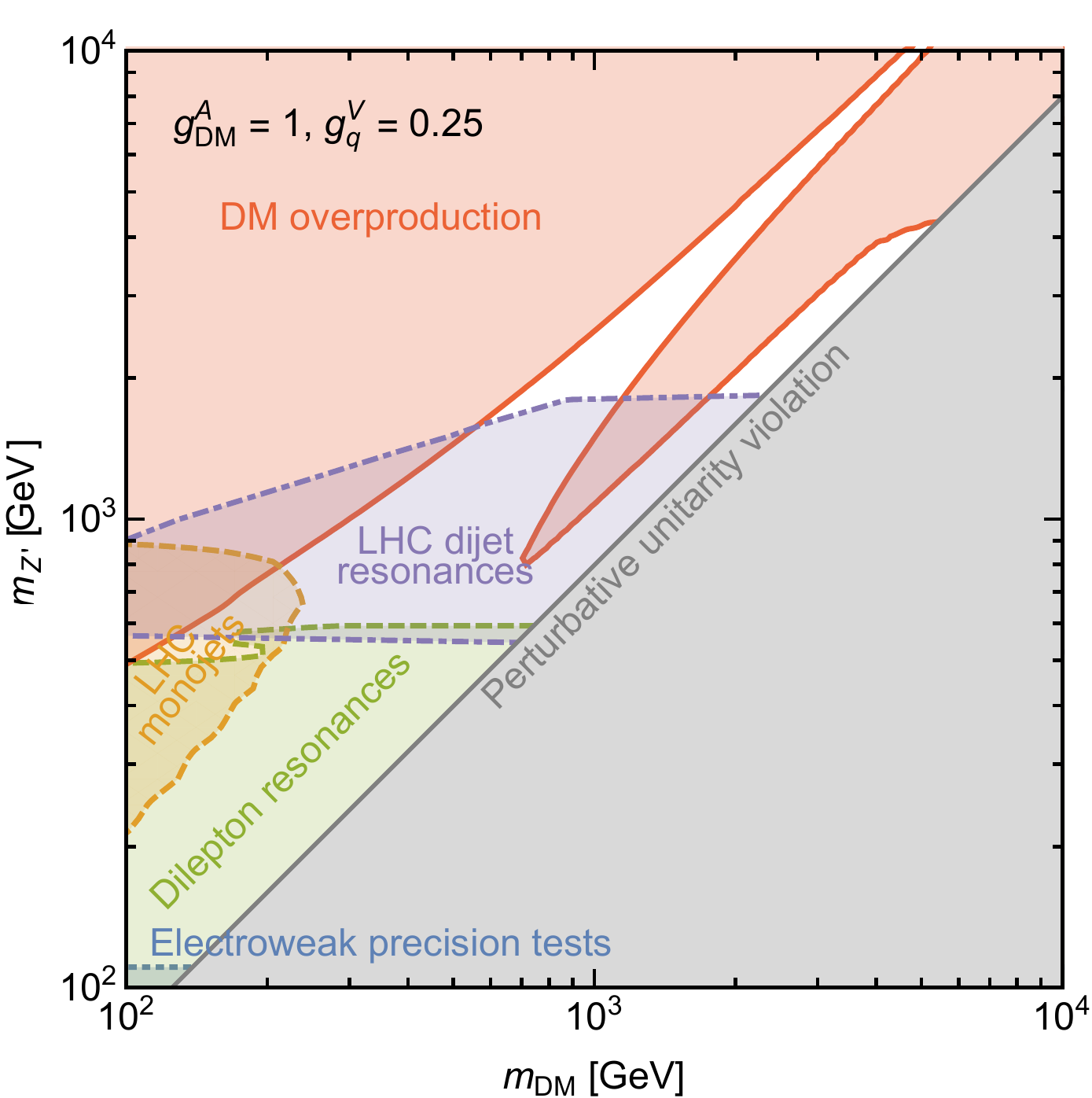}
\caption{Vector(SM)--Axial(DM): Parameter space forbidden by
  constraints from ATLAS and Tevatron dileptons (green, dashed) and
  electroweak precision observables (blue, dotted) in the
  $m_{Z'}$-$g^V_q$ parameter plane (left) and the
  $m_\text{DM}$-$m_{Z'}$ parameter plane (right), assuming that
  $\epsilon=0$ at $\Lambda = 10\:\text{TeV}$ so that kinetic mixing is
  only induced at the one-loop level.  In the right panel we
    show also the region excluded by LHC monojet (orange, dashed) and
    dijet (violet, dot-dashed) searches due to tree-level $Z'$
    exchange for the adopted coupling choice. In both panels we show
    the parameter space where perturbative unitarity is violated
    (grey). For the relic density calculation we have assumed that the
  mass of the hidden sector Higgs saturates the unitarity bound.  }
\label{fig:va_loop}
\end{figure}

While tree-level mixing is tightly constrained, it is reasonable to expect that $\epsilon$ vanishes at high scales, for example if both $U(1)$s originate from the same underlying non-Abelian gauge group, as in Grand Unified Theories. Since quarks carry charge under both $U(1)'$ and $U(1)_Y$, quark loops will still induce kinetic mixing at lower scales~\cite{Holdom:1985ag}, but the magnitude of $\epsilon$ can be much smaller than what we considered above. The precise magnitude of the kinetic mixing depends on the underlying theory, but if we assume that $\epsilon(\Lambda) = 0$ at some scale $\Lambda \gg 1\:\text{TeV}$, the kinetic mixing at a lower scale $\mu > m_t$ will be given by~\cite{Carone:1995pu}
\begin{equation}
\epsilon(\mu) = \frac{e \, g^V_q}{2 \pi^2 \, \cos \theta_\text{W}} \log \frac{\Lambda}{\mu} \simeq   0.02 \, g^V_q \, \log \frac{\Lambda}{\mu} \; .
\end{equation}
We can use this equation (setting $\mu = m_{Z'}$) to translate the
bounds from figure~\ref{fig:va} into constraints on
$g^V_q$. The results of such an analysis are shown in
  figure~\ref{fig:va_loop} assuming $\Lambda = 10$~TeV.  As can be seen
in figure~\ref{fig:va_loop} (left), searches for dilepton
resonances give again stringent constraints, implying $g_q^V <
0.1$ for $m_{Z'} = 200\:\text{GeV}$ and $g_q^V < 1$ for $m_{Z'} = 1\:\text{TeV}$.

In the right panel of figure~\ref{fig:va_loop} we also show the constraints coming from LHC searches for monojets (i.e.\ jets in association with large amounts of missing transverse energy) and for dijet resonances, adopted from ref.~\cite{Chala:2015ama}.\footnote{Note that as long as the mediator is produced on-shell, the production cross section is proportional to $(g^V_q)^2 + (g^A_q)^2$ and hence it is a good approximation to apply the bounds obtained for $g^V_q = 0$ and $g^A_q \neq 0$ also to the case $g^A_q = 0$ and $g^V_q \neq 0$. However, ref.~\cite{Chala:2015ama} assumes a Dirac DM particle, while we focus on Majorana DM. As a result, the invisible branching fraction will be somewhat smaller in our case and bounds from dijet resonance searches will be strengthened. The dijet bounds we show are therefore conservative.}
These limits are independent of the kinetic mixing $\epsilon$ since they originate from the tree-level
  $Z'$ exchange and probe larger values of $m_{Z'}$ and smaller values of $m_\text{DM}$. 
  Nevertheless, dilepton resonance searches and EWPT give relevant constraints for small $m_{Z'}$ and large $m_\text{DM}$, which are difficult to probe with monojet and dijet searches.

We conclude from figure~\ref{fig:va_loop} that the combination of constraints due to loop-induced kinetic mixing and bounds from LHC DM searches leave only a small region in parameter space (a small strip close to the $Z'$ resonance), where DM overabundance is avoided. While this result depends somewhat on our choice $\Lambda = 10$~TeV and $\mu = m_{Z'}$, it is only logarithmically sensitive to these choices.

\begin{figure}[tb]
\centering
\includegraphics[height=0.3\textheight]{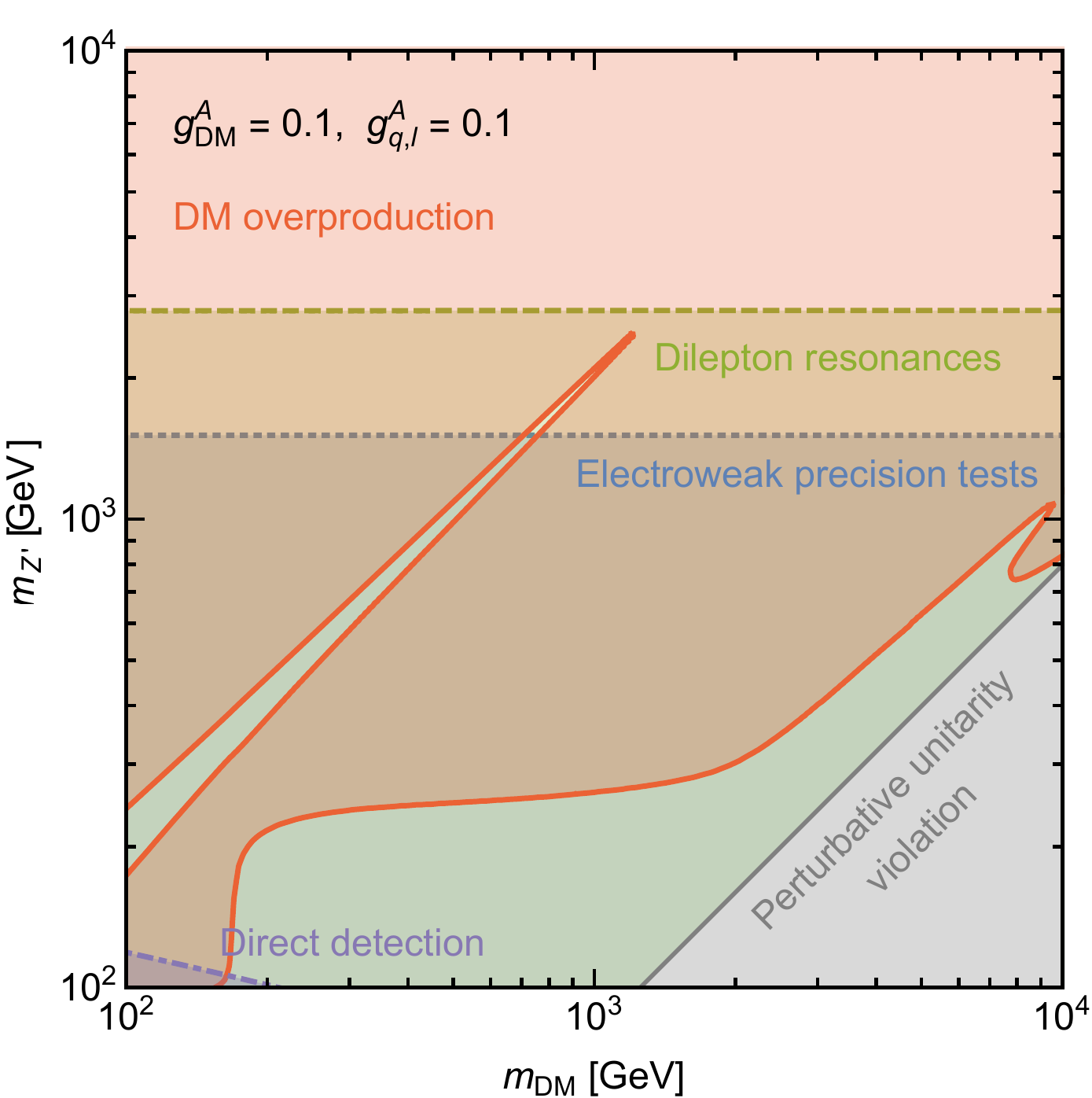}\qquad
\includegraphics[height=0.3\textheight]{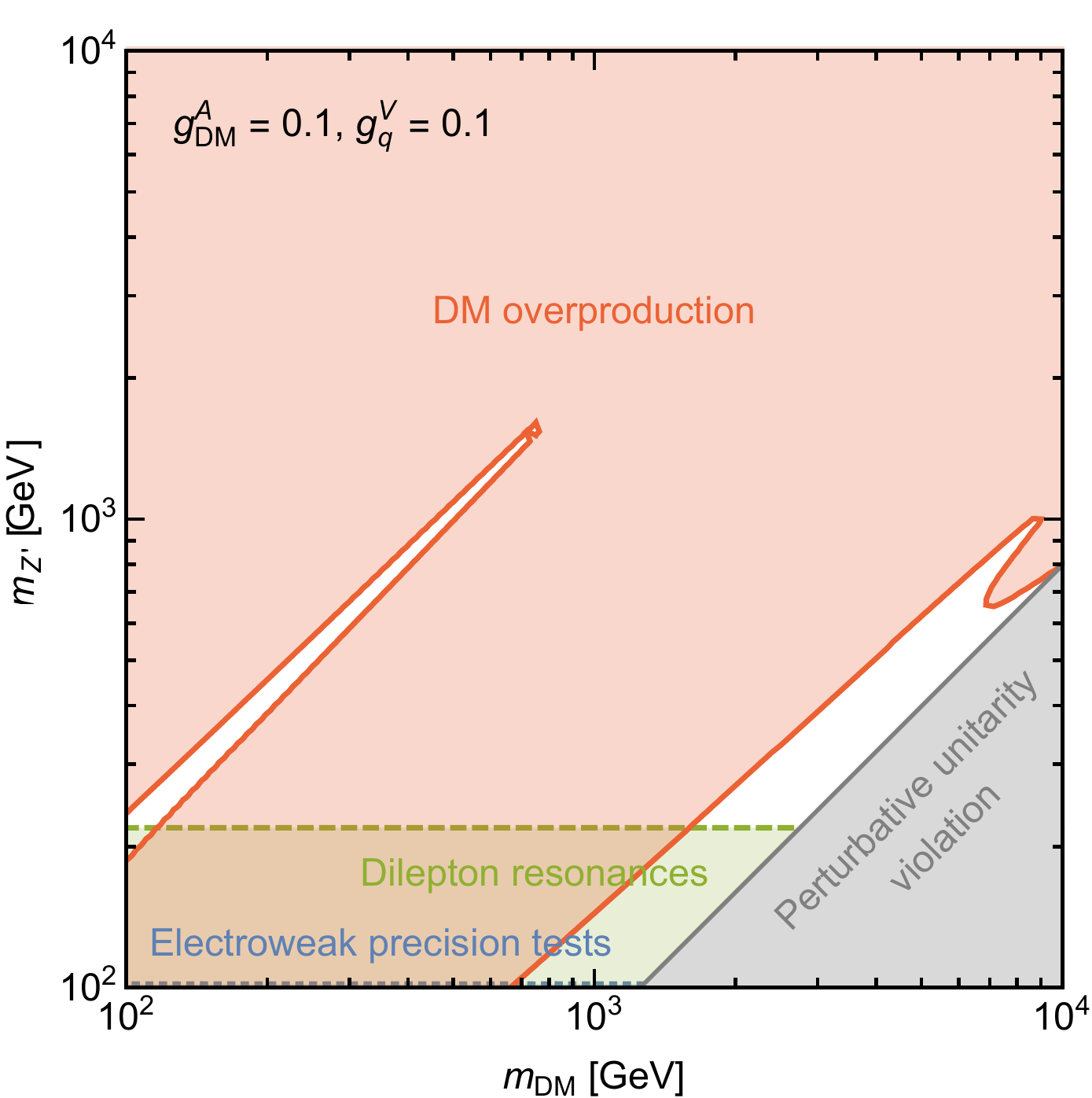}
\caption{Constraints for small DM couplings ($g^A_\text{DM} = 0.1$). The left panel considers the case Axial+Vector(SM)--Axial(DM) and should be compared to figure~\ref{fig:axialaxial2}. The right panel considers the case Vector(SM)--Axial(DM) with loop-induced kinetic mixing, assuming that $\epsilon=0$ at $\Lambda = 10\:\text{TeV}$ (cf.\ figure~\ref{fig:va_loop}).}
\label{fig:smallgA}
\end{figure}

It is worth emphasising that the unitarity constraints shown in figures~\ref{fig:axialaxial2}--~\ref{fig:va_loop} depend sensitively on the choice of $g^A_\text{DM}$ (cf.~figure~\ref{fig:ubound}). We therefore show in figure~\ref{fig:smallgA} how these constraints change if we take $g^A_\text{DM} = 0.1$ rather than $g^A_\text{DM} = 1$. In this case both $m_\text{DM}$ and $m_s$ can be much larger than $m_{Z'}$. At the same time, however, relic density constraints become significantly more severe, excluding almost the entire parameter space with $m_\text{DM} < m_{Z'}$ (apart from the resonance region $m_\text{DM} \sim m_{Z'}$). Even for $m_\text{DM} > m_{Z'}$ is it difficult to reproduce the observed relic abundance, because the annihilation channel $\chi \chi \rightarrow Z' Z'$ is significantly suppressed due to the smallness of $g^A_\text{DM}$. It only becomes relevant close to the unitarity bound, where also the process $\chi \chi \rightarrow Z Z'$ mediated by the dark Higgs gives a sizeable contribution. While in the case with axial couplings on the SM side the region compatible with thermal freeze-out becomes fully excluded by dilepton resonance searches, the case with vector couplings on the SM side is very difficult to probe at colliders and direct detection, leading to a small allowed parameter region close to the bound from perturbative unitarity.

\bigskip

In addition to the effects of kinetic mixing, we have shown above that for $g^A_\text{DM} \neq 0$ the dark sector necessarily contains a new Higgs particle. The presence of this additional Higgs can change the phenomenology of the model in two important ways. First, loop-induced couplings of the dark Higgs to SM states may give an important contribution to direct detection signals. And second, there may be mixing between the SM Higgs and the dark Higgs, leading to pertinent modifications of the properties of the SM Higgs as well as opening another portal for DM-SM interactions. We will discuss loop-induced couplings in this section and then return to a detailed study of the Higgs potential in the next section.

For $g^A_q = g^V_\text{DM} = 0$, scattering in direct detection experiments is momentum-suppressed in the non-relativistic limit and the corresponding event rates are very small. This conclusion may change if loop corrections induce unsuppressed scattering~\cite{Haisch:2013uaa}. Indeed, at the one-loop level the dark Higgs can couple to quarks and can therefore mediate unsuppressed spin-independent interactions. The resulting interaction can be written as $\mathcal{L} \propto \sum_q \, m_q \, s \, \bar{q} q$. After integrating out heavy-quark loops as well as the dark Higgs this interaction leads to an effective coupling between DM and nucleons of the form $\mathcal{L} \propto f_N \, m_N \, m_\text{DM} \, \bar{N}N \, \bar{\psi}\psi$, where $m_N$ is the nucleon mass, $N = p, n$ and $f_N \approx 0.3$ is the effective nucleon coupling.

\begin{figure}[tb]
\centering
\includegraphics[width=0.75\textwidth,clip,trim=25 670 100 15]{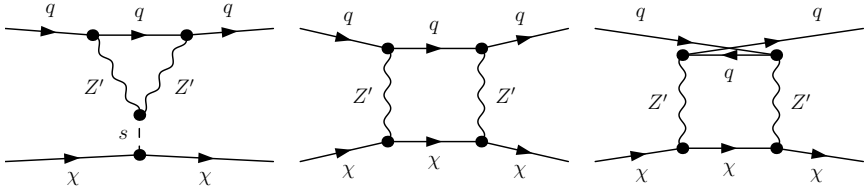}
\caption{Contributions to spin-independent scattering.}
\label{fig:SI}
\end{figure}

In the non-relativistic limit, the diagram in the left of figure~\ref{fig:SI} induces the effective interaction
\begin{equation}
\mathcal{L}_\text{eff} \supset \frac{(g^A_\text{DM})^2 \, (g^V_q)^2}{\pi^2} \frac{1}{m_s^2 \, m_{Z'}^2} \times m_\text{DM} \, f_N \, m_N \, \bar{N} N \, \bar{\psi} \psi \; .
\end{equation}
The corresponding spin-independent scattering cross section is given by
\begin{equation}
\sigma_N^\text{SI} = \frac{m_\text{DM}^2 \, f_N^2 \, m_N^2 \, \mu^2}{\pi} \frac{(g^A_\text{DM})^4 \, (g^V_q)^4}{\pi^4} \frac{1}{m_s^4 \, m_{Z'}^4} \; ,
\end{equation}
where $\mu$ is the DM-nucleon reduced mass. For masses of order $300\:\text{GeV}$ and couplings of order unity this expression yields $\sigma_N^\text{SI} \sim 10^{-46} \:\text{cm}^2$, which is below the current bounds from LUX but well within the potential sensitivity of XENON1T.

We note that there are two additional diagrams (shown in the second and third panel of figure~\ref{fig:SI}) that also lead to unsuppressed spin-independent scattering of DM particles~\cite{Haisch:2013uaa}. For $m_\text{DM} \gg m_N$, the resulting contribution is given by
\begin{equation}
\mathcal{L}_\text{eff} \supset \frac{(g^A_\text{DM})^2 \, (g^V_q)^2}{4 \pi^2} \frac{m_\text{DM}^2 - m_{Z'}^2 + m_{Z'}^2 \log (m_{Z'}^2 / m_\text{DM}^2)}{m_{Z'}^2 (m_{Z'}^2 - m_\text{DM}^2)^2} \times m_\text{DM} \, f_N \, m_N \, \bar{N} N \, \bar{\psi} \psi \; .
\end{equation}
If there is no large hierarchy between $m_\text{DM}$, $m_{Z'}$ and $m_s$, this contribution is of comparable magnitude to the one from dark Higgs exchange and interference effects can be important. Moreover, there may be a relevant contribution from loop-induced spin-dependent scattering. We leave a detailed study of these effects to future work.

\section{Mixing between the two Higgs bosons}
\label{sec:higgsmixing}

In addition to the loop-induced couplings of the dark Higgs to SM fermions discussed in the previous
section, such couplings can also arise at tree-level from mixing. In fact, an important
implication of the presence of a second Higgs field is that the two
Higgs fields will in general mix, thus modifying the properties of the
mostly SM-like Higgs. Furthermore, the mixing opens up the so-called Higgs portal between the
DM and SM particles, leading to a much richer DM phenomenology than in
the case of DM-SM interactions only via the vector mediator.

The mixing between the scalars is due to an additional term in the scalar potential:
\begin{equation}
  V(S, H) \supset \lambda_{hs} (S^*S)(H^\dagger H) \,.
\end{equation}
The coupling $\lambda_{hs}$ is a free parameter, independent of the vector mediator. For non-zero $\lambda_{hs}$, the scalar mass eigenstates $H_{1,2}$ are given by
\begin{align}
 H_1 = s \sin \theta + h \cos \theta \nonumber \\
 H_2 = s \cos \theta - h \sin \theta 
\end{align}
where, as shown in App.~\ref{app:scalar},
\begin{equation}
 \theta \approx - \frac{\lambda_{hs} \, v \, w}{m_s^2 - m_h^2} + \mathcal{O}(\lambda_{hs}^3) \; .
\end{equation}
We emphasise  that perturbative unitarity implies that $m_s$ cannot be arbitrarily large (for given $m_{Z'}$ and $g^A_\text{DM}$) and hence it is impossible to completely decouple the dark Higgs.

The resulting Higgs mixing leads to three important consequences. First, the (mostly) dark Higgs obtains couplings to SM particles, enabling us to produce it at hadron colliders and to search for its decay products (or monojet signals). Second, the properties of the (mostly) SM-like Higgs, in particular its total production cross section and potentially also its branching ratios, are modified. And finally, both Higgs particles can mediate interactions between DM and nuclei, leading to potentially observable signals at direct detection experiments.

\begin{figure}[tb]
\centering
\includegraphics[height=0.3\textheight]{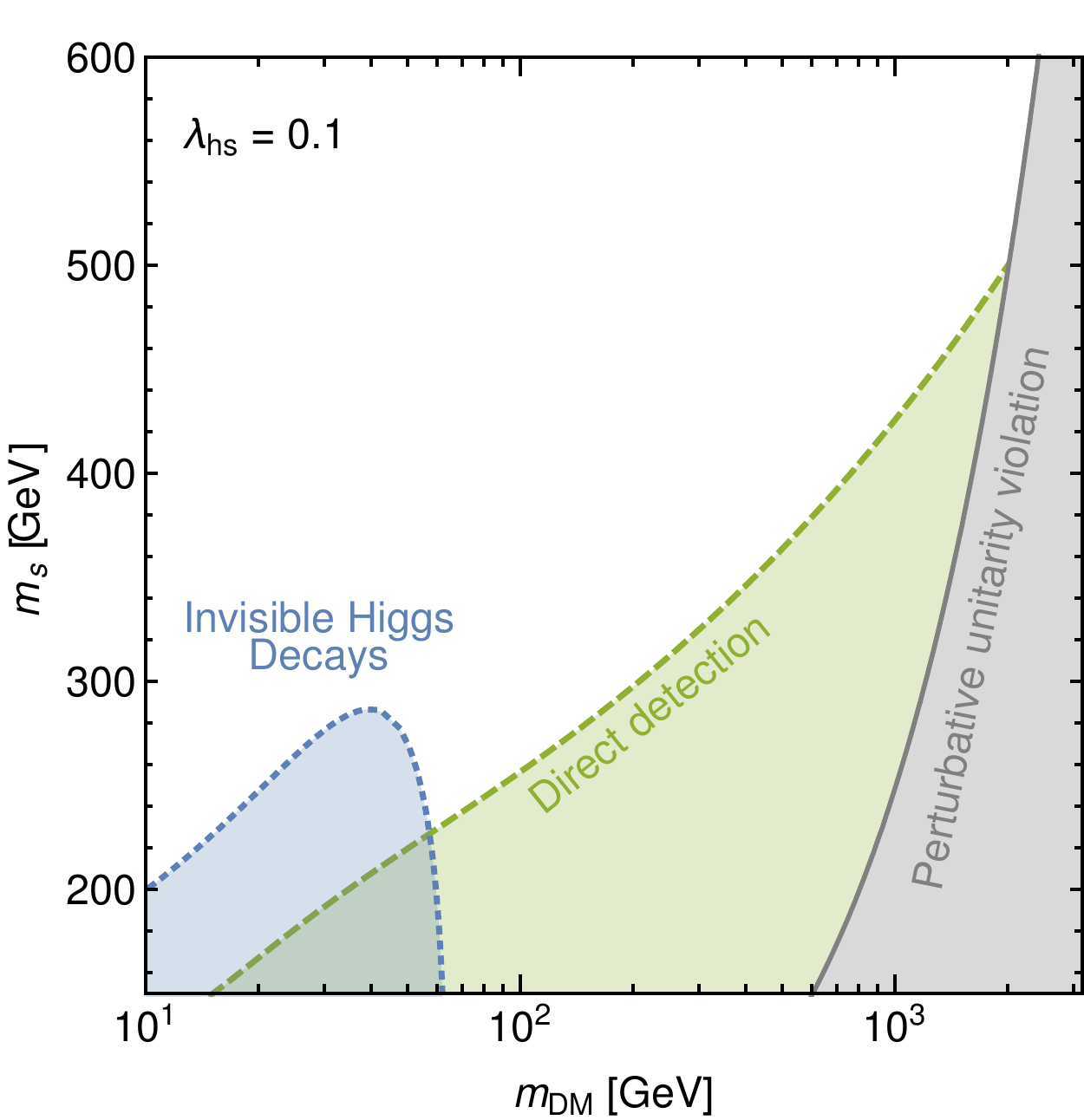}\qquad\includegraphics[height=0.3\textheight]{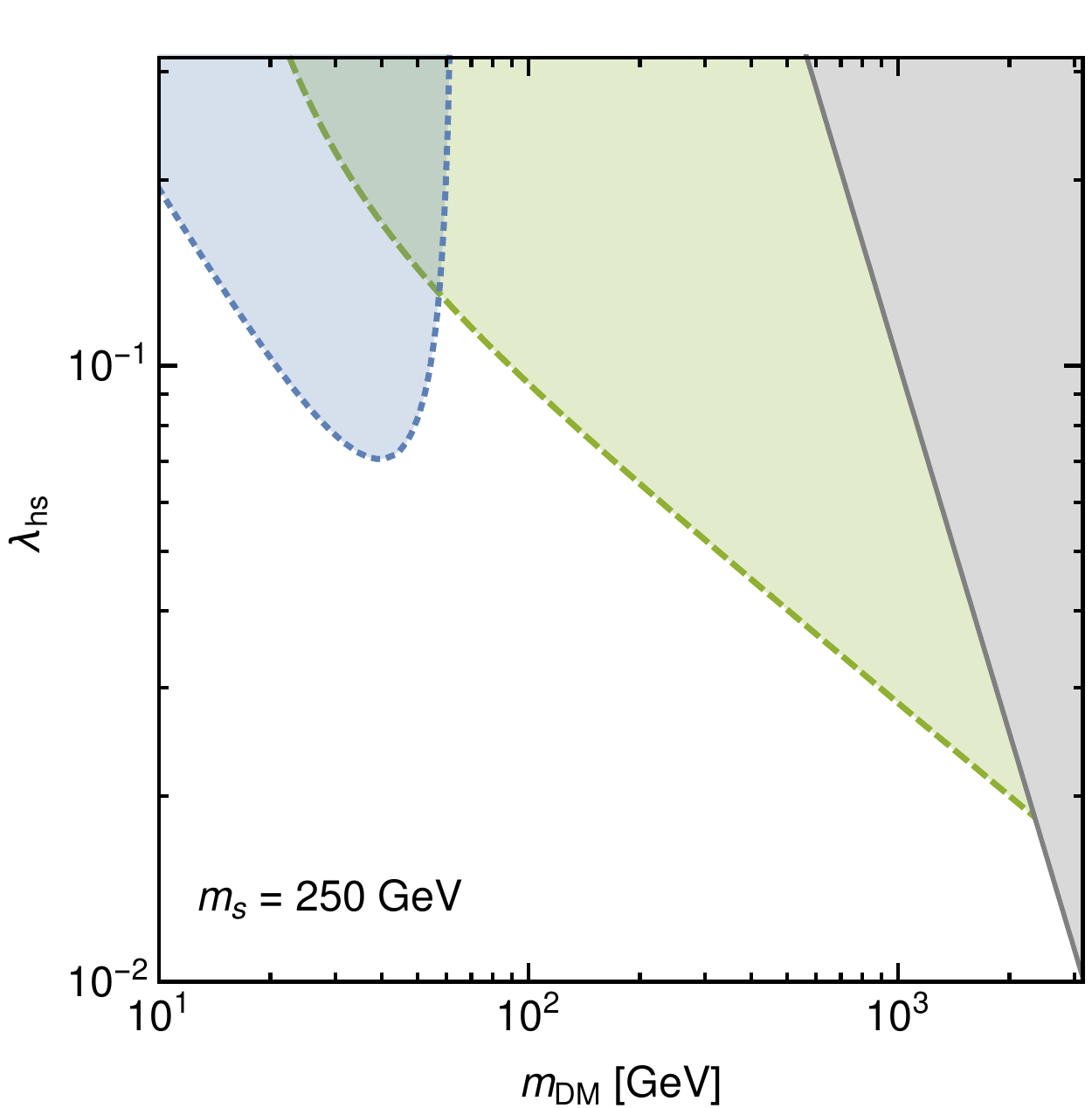}
\caption{Constraints on $m_s$ and $\lambda_{hs}$ from bounds on the Higgs invisible branching ratio (blue, dotted) and from bounds on the spin-independent DM-nucleon scattering cross section (green, dashed). In the grey parameter region unitarity constraints are in conflict with the stability of the potential.}
\label{fig:higgsmixing}
\end{figure}

Higgs portal DM has been extensively studied, see for instance
\cite{Djouadi:2011aa, Lebedev:2012zw, LopezHonorez:2012kv,
  Baek:2012uj,  Walker:2013hka, Esch:2013rta, Freitas:2015hsa} for an incomplete
selection of references.  A full analysis of Higgs mixing effects is beyond
the scope of the present paper. Nevertheless, to illustrate the
magnitude of potential effects, let us consider the induced coupling
of the SM-like Higgs $H_1 \approx h$ to DM particles
\begin{equation}
\mathcal{L} \supset - \frac{m_\text{DM} \, \sin \theta}{2 \, w} h \, \bar{\psi} \psi \simeq \frac{m_\text{DM} \, \lambda_{hs} \, v}{2 (m_s^2 - m_h^2)} h \, \bar{\psi} \psi \; . 
\end{equation}
For small $\lambda_{hs}$, the resulting direct detection cross section is given by~\cite{Djouadi:2011aa}
\begin{equation}
 \sigma_N^\text{SI} \simeq \frac{\mu^2}{\pi \, m_h^4} \frac{f_N^2 \, m_N^2 \, m_\text{DM}^2\,\lambda_{hs}^2}{(m_s^2 - m_h^2)^2} \; ,
\end{equation}
where we can neglect an additional contribution from the exchange of a dark Higgs provided $m_s^4 \gg m_h^4$.
The parameter regions excluded by the LUX results~\cite{Akerib:2013tjd} are shown in figure~\ref{fig:higgsmixing} (green regions).

We note that (in the linear approximation) the direct detection cross section is independent of $w$ and does therefore not depend on $m_{Z'}$ or $g^A_\text{DM}$. Nevertheless $w$ is not arbitrary, because unitarity gives a lower bound $\sqrt{4\pi} w > \text{max}\left[\sqrt{2} m_\text{DM}, m_s \right]$. At the same time, stability of the Higgs potential requires $4 \lambda_s \, \lambda_h > \lambda_{hs}$. These two inequalities can only be satisfied at the same time if
\begin{equation}
m_\text{DM} < \frac{\sqrt{2 \pi} \, m_s \, m_h}{v} \; .
\end{equation}
In figure~\ref{fig:higgsmixing}, we show the parameter region where unitarity and stability are in conflict in grey.

Finally, if the DM mass is sufficiently small, the SM-like Higgs can decay into pairs of DM particles, with a partial width given by~\cite{Djouadi:2011aa}
\begin{equation}
 \Gamma^\text{inv} = \frac{1}{8 \pi} \frac{m_\text{DM}^2 \, \lambda_{hs}^2}{(m_s^2 - m_h^2)^2} v^2 m_h \left(1-\frac{4 \, m_\text{DM}^2}{m_h^2}\right)^{3/2} \; .
\end{equation}
The invisible branching fraction is tightly constrained by LHC measurements: $\text{BR}(h\rightarrow \text{inv}) < 0.27$~\cite{Khachatryan:2014jba}. Furthermore, a combined fit from ATLAS and CMS yields $\mu = 1.09^{+0.11}_{-0.10}$ for the total Higgs signal strength~\cite{ATLAS-CONF-2015-044}, which can be used to deduce $\text{BR}(h\rightarrow \text{inv}) < 0.11$ at 95\% CL. The resulting constraints, compared to the ones on $\sigma_N^\text{SI}$ from LUX, are shown in figure~\ref{fig:higgsmixing} (blue regions).

The crucial observation is that the necessary presence of a dark Higgs will in general induce additional signatures and therefore lead to new ways to constrain models with a $Z'$ mediator using both direct detection experiments and Higgs measurements. However, since $\lambda_{hs}$ and $m_s$ are effectively free parameters, it is difficult to directly compare the constraints shown in figure~\ref{fig:higgsmixing} to the ones obtained from monojet and dijet searches at the LHC. Nevertheless, we can conservatively estimate the relevance of these effects by fixing the dark Higgs mass $m_s$ to the largest value consistent with perturbative unitarity. 

\begin{figure}[tb]
\centering
\includegraphics[height=0.3\textheight]{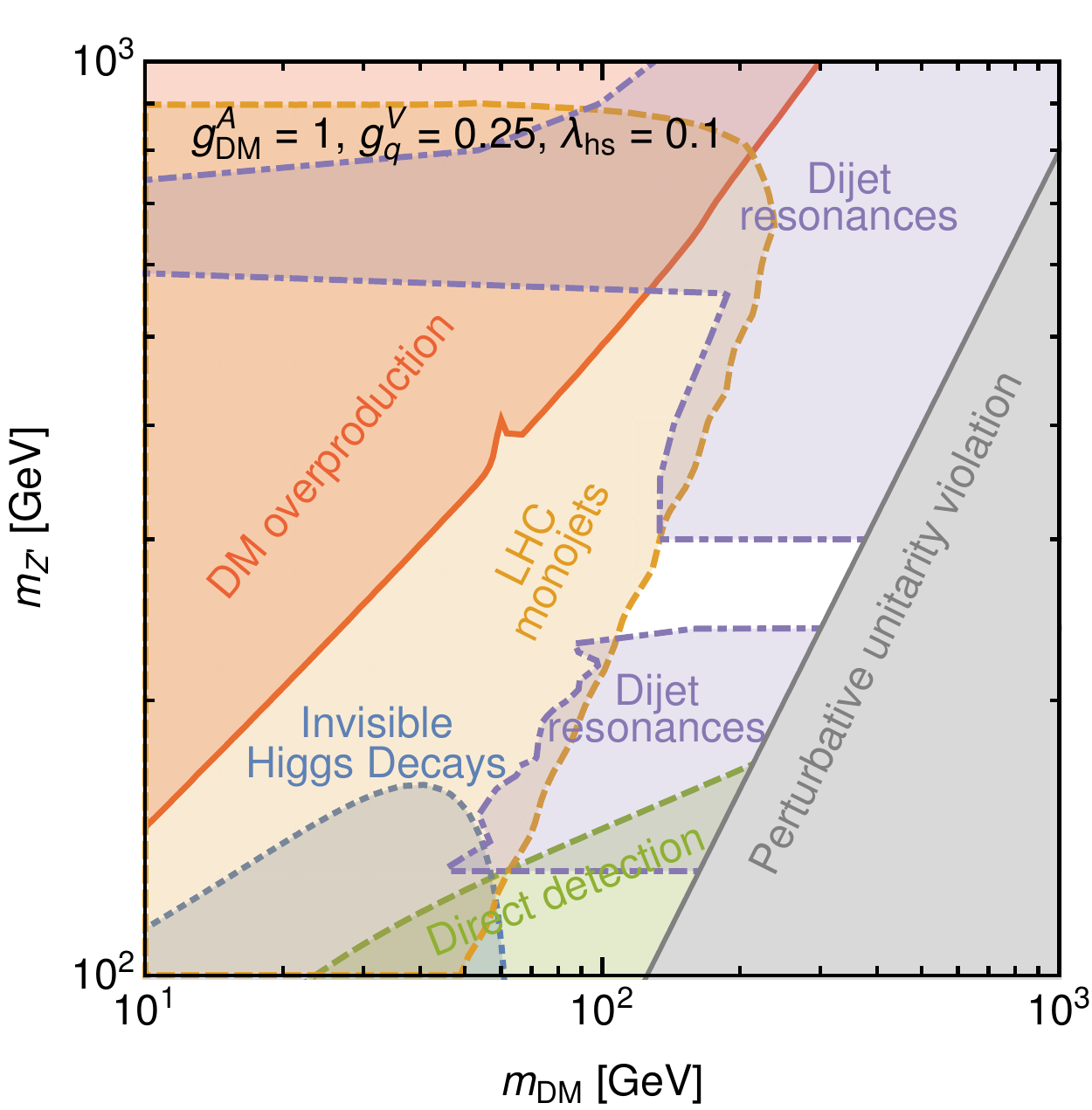}\qquad\includegraphics[height=0.3\textheight]{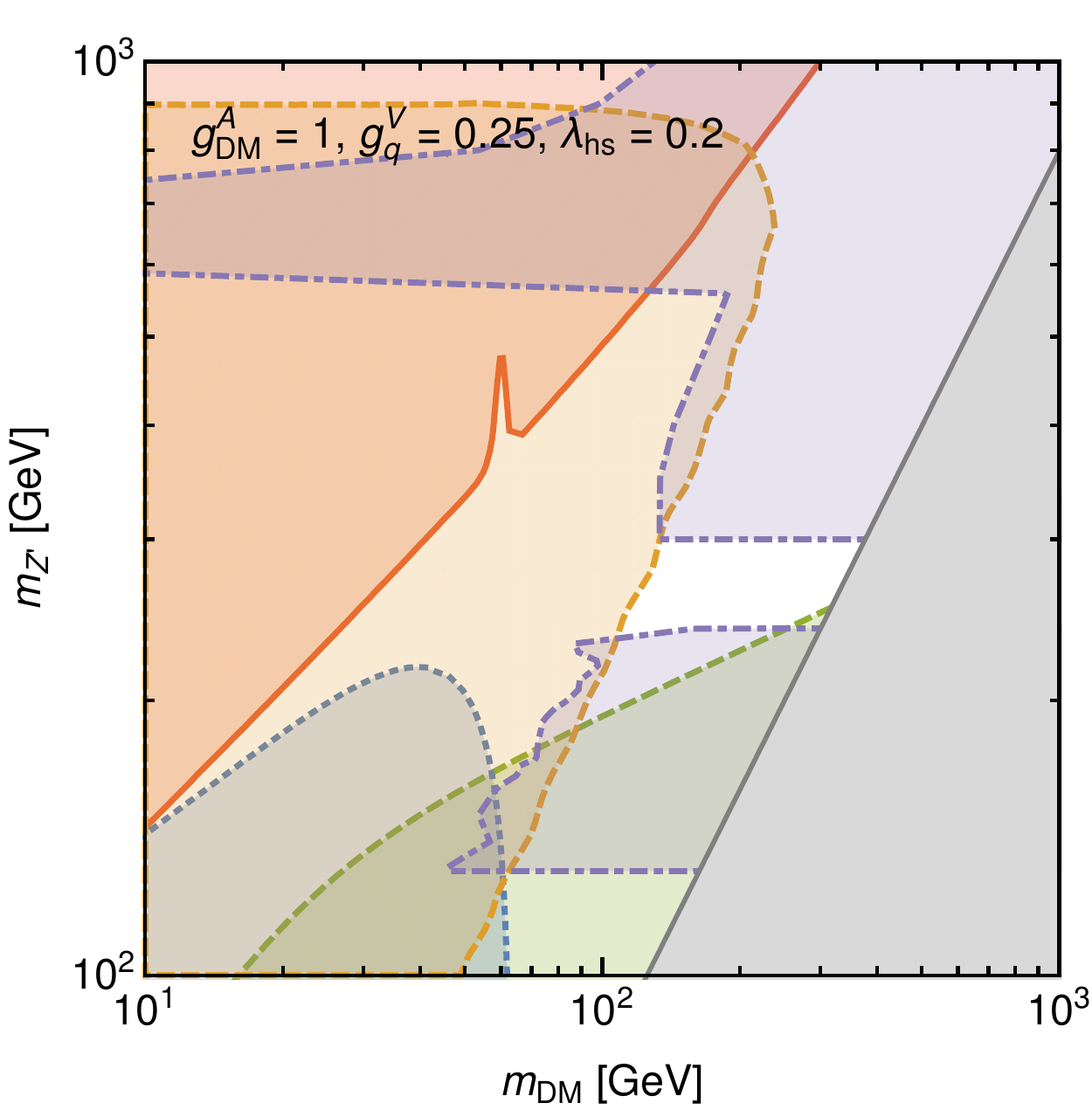}
\caption{Constraints in the $m_{Z'}$-$m_\text{DM}$ plane for different values of $\lambda_{hs}$, taking the mass of the hidden sector Higgs to saturate the unitarity bound. The blue (dotted) region is excluded by bounds on the Higgs invisible branching ratio and the green (dashed) region is in conflict with bounds on the spin-independent DM-nucleon scattering cross section. The orange (dashed) region shows constraints from the CMS monojet search, the purple (dot-dashed) region is excluded by a combination of dijet searches from the LHC, Tevatron and UA2 (adopted from ref.~\cite{Chala:2015ama}). In the grey parameter region unitarity constraints are in conflict with the stability of the potential, the red region corresponds to DM overproduction. Note the change of scale in these figures.}
\label{fig:higgsmixing2}
\end{figure}

The resulting constraints in the conventional $m_{Z'}$-$m_\text{DM}$ parameter plane with fixed couplings are shown in figure~\ref{fig:higgsmixing2}. For comparison we show the constraints from the CMS monojet search~\cite{Khachatryan:2014rra} and a combination of dijet searches from the LHC, Tevatron and UA2 (adopted from ref.~\cite{Chala:2015ama}). We find that the additional constraints due to Higgs mixing provide valuable complementary information in the parameter region with small $m_{Z'}$ and large $m_\text{DM}$, which is difficult to probe with monojet or dijet searches. Note that for $m_{Z'} > 2\:\text{TeV}$ (not shown in figure~\ref{fig:higgsmixing2}) there is still an allowed parameter region if either $m_\text{DM} \approx m_{Z'} / 2$ or $m_\text{DM} > m_{Z'}$ (cf.\ figure 6). Furthermore, it is worth emphasising that for smaller values of $m_s$ significantly stronger constraints are expected from Higgs mixing. Moreover, these constraints are independent of the SM couplings of the $Z'$ and will therefore become increasingly important in the case of small $g_q$.

\section{Discussion and Outlook}
\label{sec:discussion}

In this paper we have studied the so-called simplified model approach to DM used to parametrise the interactions of a DM
  particle with the SM via one or several new mediators. It should be clear that simplified models are
  considered merely as an effective description, used as a tool to
  combine different DM search strategies. Nevertheless, it is
  important that such models fulfil basic requirements, such as gauge invariance and that
  perturbative unitarity is guaranteed in the regions of the parameter
  space where the model is used to describe data. To ensure gauge
  invariance, one needs to impose certain relations between the
  different couplings, whereas it is necessary to introduce additional
  states in order to restore perturbative unitarity.
  
We have illustrated these issues by considering a simplified model consisting of a fermionic DM particle and a vector mediator, which may for example be the $Z'$ gauge boson of a new $U(1)'$ gauge symmetry in the hidden sector. The phenomenology of this model depends decisively on whether the couplings of the mediator are purely vectorial or whether there are non-zero axial couplings (implying that left- and right-handed fields are charged differently under the new $U(1)'$). Since the coupling structure on the SM side may be different from the one of the DM side, there are four different cases of interest: purely vectorial couplings on both sides, non-zero axial couplings on either the SM side or the DM side, and non-zero axial couplings in both sectors. Our results can be summarized as follows:

\begin{enumerate}
\item \label{it:VV} {\bf Vector(SM)--Vector(DM):} In this case no additional new physics is needed to guarantee perturbative unitarity and the mass of the $Z'$ can be generated via the Stueckelberg mechanism. This model is however highly constrained phenomenologically and a thermal DM is excluded for large parts of the parameter space due to strong limits on the spin-independent DM-nucleon scattering cross section.
\end{enumerate}

\noindent Generally, if at least one of the axial couplings is non-zero one needs new physics to unitarize the longitudinal component of the $Z'$. As a simple example we consider a SM-singlet Higgs breaking the dark $U(1)'$. Unitarity then requires the mass of the new Higgs to be comparable to the $Z'$ mass. Models with non-zero axial couplings are therefore expected to have a rich phenomenology with promising experimental signatures in DM direct detection experiments and invisible Higgs decays as well as additional DM annihilation channels.

\begin{enumerate}
\setcounter{enumi}{1}
\item \label{it:AA} {\bf Axial(SM)--Axial(DM):} The crucial observation in this case is that gauge invariance of the SM Yukawa terms requires that the SM Higgs has to be charged under the $U(1)'$. This requirement has important implications for phenomenology:
  \begin{itemize}
    \item[(a)] Electroweak symmetry breaking leads to mass mixing between the $Z'$ and the SM $Z$-boson, which is strongly constrained by EWPT.
    \item[(b)] The axial couplings of SM fermions to the $Z'$ are necessarily flavour universal and equal for quarks and leptons. Hence, it is not possible to couple the DM particle to quarks without also inducing couplings of the $Z'$ to leptons. Since the LHC is very sensitive to dilepton resonances, the resulting bounds severely constrain the model (dominating over constraints from monojet and dijet searches).
  \end{itemize}

\item \label{it:AV} {\bf Axial(SM)--Vector(DM):} The constraints from EWPT and dilepton resonance searches are largely independent of the coupling between the $Z'$ and DM and therefore also apply in the case of purely vectorial couplings on the DM side. However, gauge invariance of the Yukawa couplings implies that it is impossible for the $Z'$ to have purely axial couplings to quarks. Consequently, as soon as there is a vectorial coupling on the DM side, one necessarily obtains a vector-vector component inducing unsuppressed spin-independent DM-nucleus scattering, which is strongly constrained by direct detection (see item~\ref{it:VV} above).

\item \label{it:VA} {\bf Vector(SM)--Axial(DM):} In contrast to the couplings between the $Z'$ and quarks it is possible for the DM-$Z'$ coupling to be purely axial. Indeed, this situation arises naturally in the case that the DM particle is a Majorana fermion such that the vector current vanishes. If the couplings on the SM side are purely vectorial (i.e.\ left- and right-handed SM fields have the same charge), the SM Higgs is uncharged under the $U(1)'$ and consequently the constraints discussed in item~\ref{it:AA} do not apply. Furthermore, the tree-level direct detection cross section is velocity suppressed, leading to much weaker constraints on this particular scenario.

Nevertheless, sizeable spin-independent DM-nucleus scattering can be induced at loop level. In addition, kinetic mixing between the $Z'$ and SM gauge bosons (at tree level or loop-induced) can be potentially important for EWPT and dilepton signatures. Assuming $\epsilon = 0$ at $\Lambda=10$~TeV, we find that bounds from searches for dilepton resonances due to loop-induced kinetic mixing can still be relevant and give constraints that are complementary to the ones obtained from monojet and dijet searches.

\end{enumerate}

All in all we find that imposing gauge invariance and conservation of perturbative unitarity has important implications for the phenomenology of DM interacting via a vector mediator and that relevant experimental signatures are not captured by considering only the interactions of the vector mediator with DM and quarks. This observation is relevant for the interpretation of various recent analyses of $Z'$-based simplified models,  e.g.~\cite{Lebedev:2014bba,Hooper:2014fda,Chala:2015ama,Alves:2015dya,Blennow:2015gta,Heisig:2015ira}. Indeed, the $Z'$ model considered here is severely constrained by EWPT and dilepton resonance searches, either due to tree-level effects or loop-induced kinetic mixing. Moreover, the general expectation is that the mixing between the dark Higgs and the SM Higgs is sizeable and that as a result Higgs portal interactions are present in addition to the interactions mediated by the $Z'$. 

The weakest constraints are obtained in the case of purely vectorial couplings on the SM side and purely axial couplings on the DM side. Indeed, this is the only case where LHC monojet and dijet searches are potentially competitive with other kinds of constraints.   In all cases that we have considered we find the hypothesis of thermal DM production to be under significant pressure. In large regions of the  parameter space which is still allowed by experiments additional annihilation channels (beyond the $Z'$-mediated interactions) are necessary to avoid DM
overabundance. A more systematic parameter scan of the model will be performed in a forthcoming publication~\cite{upcoming}. 

Two final comments are in order. First, in this work we have not taken into account gauge anomalies. In general new fermions are needed to cancel the anomalous triangle diagrams, potentially leading to additional signatures and further constraints on the model.
However, due to the constraints implied by gauge invariance of the SM Yukawa terms, the gluon-gluon-$Z'$ anomaly vanishes automatically, so that the new fermions need not be charged under colour, making them difficult to probe at the LHC.

Second, the requirement of universality of all axial fermion charges (including leptons) follows from the gauge invariance of the SM Yukawa term. It relies on the fact that in the SM all fermion masses are generated by the same Higgs doublet. If the Higgs sector is more complicated, for example in a two-Higgs-doublet model, this condition is relaxed and it is possible to have different axial couplings to up- and down-type quarks or different axial couplings to quarks and to leptons. In any case, such extensions of the SM go significantly beyond the simplified model approach and would most likely have a number of implications for Higgs physics, EWPT and other searches for new physics.

\bigskip

In conclusion we would like to emphasise that one of the most intriguing implications of our study is that a model with a vector mediator should generically also contain a scalar mediator, corresponding to the dark Higgs that generates the vector mass. In the limit that the mass of the vector mediator is much larger than the mass of the scalar, our $Z'$ model can also be used to study simplified models with a scalar mediator, where gauge invariance and perturbative unitarity can be similarly problematic. Indeed, our findings suggest that a strict distinction between simplified models with scalar and vector mediators is unnatural and many of the issues with these models may be best addressed in a more realistic set-up combining the two. Future direct detection experiments together with the upcoming runs of the LHC will be able to thoroughly explore the parameter space of such a realistic simplified model and test the hypothesis of DM as a thermal relic.

\subsection*{Acknowledgements}

We thank Sonia El Hedri, Juan Herrero-Garcia, Matthew McCullough and Oscar St\aa l for helpful discussions, and Michael Duerr and Jure Zupan for valuable comments on the manuscript. FK would like to thank the Oskar Klein Center and Stockholm University for hospitality. This work is supported by the German Science Foundation (DFG) under the Collaborative Research Center (SFB) 676 Particles, Strings and the Early Universe as well as the ERC Starting Grant `NewAve' (638528).

\begin{appendix}
\section{Coupling structure from mixing}
\label{ap:mixing}

\subsection{Gauge boson mixing}
\label{app:Z}

In this appendix we discuss the mixing of a gauge boson $\hat{Z}'$ of a new $U(1)'$ gauge group with the SM $U(1)_Y$ gauge field $\hat B$ and the neutral component $\hat W^3$ of the $SU(2)_\mathrm{L}$ weak fields, where we use hats to denote the interaction eigenstates in the original basis. The mixing will then lead to the mass eigenstates $Z'$, $Z$ and $A$. Following the discussion in~\cite{Frandsen:2011cg}, we consider an effective Lagrangian including both kinetic mixing and mass mixing (see also~\cite{Babu:1997st})
\begin{align}
  {\cal L} =& \; {\cal L}_{SM}
  -\frac{1}{4}\hat{X}^{\mu\nu}\hat{X}_{\mu\nu} + {\frac{1}{2}} m_{\hat
    Z'}^2 \hat{Z'}_\mu \hat{Z'}^\mu - {\frac{1}{2}} \sin \epsilon\, \hat{B}_{\mu\nu} \hat{X}^{\mu\nu} +\delta
  m^2 \hat{Z}_\mu \hat{Z'}^\mu \;,
\label{eq:Lappendix}
\end{align}
where $\hat{X}^{\mu \nu} \equiv \partial^\mu \hat{Z'}^\nu - \partial^\nu \hat{Z'}^\mu$. Furthermore, we have defined $\hat{Z}\equiv \hat{c}_\mathrm{W} \hat{W}^3- \hat{s}_\mathrm{W} \hat{B}$, where $\hat{s}_\mathrm{W} \, (\hat{c}_\mathrm{W})$ is the sine (cosine) of the Weinberg angle and $\hat g', \, \hat g$ are the corresponding gauge couplings.

The field strengths are diagonalised and canonically normalised via the following two consecutive transformations~\cite{Babu:1997st,Chun:2010ve,Frandsen:2011cg}
\begin{align}
\label{eq:Zpmixing}
\left(\begin{array}{c} \hat B_\mu \\ \hat W_\mu^3 \\ \hat
    Z'_\mu \end{array}\right) & = \left(\begin{array}{ccc} 1 & 0 &
    -t_\epsilon \\ 0 & 1 & 0 \\ 0 & 0 &
    1/c_\epsilon \end{array}\right)
\left(\begin{array}{c} B_\mu \\ W_\mu^3 \\ Z'_\mu \end{array}\right) \ , \\
\left(\begin{array}{c} B_\mu \\ W_\mu^3 \\ Z'_\mu \end{array}\right) &
= \left(\begin{array}{ccc}
    \hat c_\mathrm{W} & -\hat s_\mathrm{W} c_\xi &  \hat s_\mathrm{W} s_\xi \\
    \hat s_\mathrm{W} & \hat c_\mathrm{W} c_\xi & - \hat c_\mathrm{W} s_\xi \\
    0 & s_\xi & c_\xi
\end{array} \right) 
\left(\begin{array}{c} A_\mu \\ Z_\mu \\ R_\mu \end{array}\right)
 \; ,
\end{align}
where
\begin{align}
  t_{2\xi}=\frac{-2c_\epsilon(\delta m^2+m_{\hat Z}^2 \hat
    s_\mathrm{W} s_\epsilon)} {m_{\hat Z'}^2-m_{\hat
      Z}^2 c_\epsilon^2 +m_{\hat Z}^2\hat s_\mathrm{W}^2 s_\epsilon^2
    +2\,\delta m^2\,\hat s_\mathrm{W} s_\epsilon} \; .
\label{eq:xi}
\end{align}
For $\epsilon \ll 1$ and $\delta m^2 \ll m_{\hat Z}^2, m_{\hat Z'}^2$, this equation can be approximated by
\begin{equation}\label{eq:xi_def}
 \xi = \frac{\delta m^2 + m_{\hat Z}^2 \hat s_\mathrm{W} \epsilon}{m_{\hat Z}^2 - m_{\hat Z'}^2} \; .
\end{equation}

The mass eigenvalues $m_Z$ and $m_{Z'}$ are given by
\begin{align} 
 m_Z^2 & = m_{\hat Z}^2 (1+{\hat s}_\mathrm{W} \, t_\xi \, t_\epsilon)+\frac{\delta m^2 \, t_\xi}{c_\epsilon} \nonumber \\ & \approx m_{\hat Z}^2 + (m_{\hat Z}^2 - m_{\hat Z'}^2) \xi^2 \; ,  \\
 m_{Z'}^2 & = \frac{m_{\hat Z'}^2 + \delta m^2 ({\hat s}_\mathrm{W} \, s_\epsilon-c_\epsilon \, t_\xi)}{c_\epsilon^2 \, (1+{\hat s}_\mathrm{W} \, t_\xi \, t_\epsilon)} \nonumber \\ &  \approx m_{\hat Z'}^2 + m_{\hat Z'}^2 \xi (\xi - {\hat s_\mathrm{W}} \epsilon) - m_{\hat Z}^2 (\xi - {\hat s_\mathrm{W}} \epsilon)^2\; .
\end{align}
We define the `physical' weak angle via
\begin{align}
  s_\mathrm{W}^2 \, c_\mathrm{W}^2=\frac{\pi \, \alpha(m_{Z})}{\sqrt{2} \, G_\mathrm{F} \, m_{Z}^2} \; ,
\label{eq:swcw}
\end{align}
where $\alpha = e^2 / (4\pi)$. Eq.~(\ref{eq:swcw}) also holds with the replacements $s_\mathrm{W}\to\hat s_\mathrm{W}$, $c_\mathrm{W}\to\hat c_\mathrm{W}$ and $m_{Z}\to m_{\hat Z}$, leading to the identity $s_\mathrm{W} \, c_\mathrm{W} \, m_{Z}=\hat s_\mathrm{W} \, \hat c_\mathrm{W} \, m_{\hat Z}$.  This equation implies
\begin{equation}
 s_\mathrm{W}^2 = \hat s_\mathrm{W}^2 - \frac{\hat s_\mathrm{W}^2 \, \hat c_\mathrm{W}^2}{\hat c_\mathrm{W}^2 - \hat s_\mathrm{W}^2} \left(1 - \frac{m_{\hat Z'}^2}{m_{\hat Z}^2}\right) \xi^2 \; .
\end{equation}
These equations allow us to fix $\hat s_\mathrm{W}$ and $m_{\hat Z}$ in such a way that we reproduce the experimentally well-measured quantities $s_\mathrm{W}$ and $m_{Z}$.

The couplings of the $Z'$ to SM fermions induced via mixing can e.g.\ be found in \cite{Frandsen:2012rk}.
Of particular interest to our current analysis are the couplings to leptons which are strongly constrained.
In terms of the mixing parameters they can be written as
\begin{align}
  g_{\ell}^\mathrm{V} &= \frac{1}{4}(3 {\hat g'} (\hat s_\mathrm{W} s_\xi-c_\xi t_\epsilon)- {\hat
    g}\hat c_\mathrm{W} s_\xi) \ , & g_{\ell}^\mathrm{A} &= -\frac{1}{4}
  ({\hat g'} (\hat s_\mathrm{W} s_\xi-c_\xi t_\epsilon) + {\hat g} \hat c_\mathrm{W} s_\xi) \; ,
\label{eq:dilepton}
\end{align}
with $\hat{g}$ and $\hat{g}'$ the fundamental gauge couplings of 
$SU(2)_\text{L}$ and $U(1)_Y$.

\subsection{Scalar mixing}
\label{app:scalar}

Considering the SM Higgs $h$ plus the dark Higgs $s$, the most general
scalar potential after electroweak and dark symmetry breaking can be
written as
\begin{equation}
 V(s, h) = - \frac{\mu_s^2}{2} (s+w)^2  - \frac{\mu_h^2}{2} (h+v)^2 + \frac{\lambda_h}{4} (h+v)^4 + \frac{\lambda_s}{4} (s+w)^4 + \frac{\lambda_{hs}}{4} (h+v)^2(s+w)^2 \; .
 \end{equation}
For $\lambda_{hs} = 0$, we obtain the usual formulas
\begin{align}
 v^2 & = \frac{\mu_h^2}{\lambda_h} \; , \quad m_h^2 = 2 \, \lambda_h \, v^2 \; ,\\
 w^2 & = \frac{\mu_s^2}{\lambda_s} \; , \quad m_s^2 = 2 \, \lambda_s \, w^2 \; .
\end{align}
In this case, there is no mixing between the two Higgs fields even at one-loop level. Nevertheless, there is no reason why $\lambda_{hs}$ should be negligible and therefore the two fields will in general mix. One then obtains for the minimum (assuming $4 \, \lambda_h \, \lambda_s > \lambda_{hs}^2$)
\begin{align}
 v^2 & = 2 \frac{2 \, \lambda_s \, \mu_h^2 - \lambda_{hs} \, \mu_s^2}{4 \, \lambda_s \, \lambda_h - \lambda_{hs}^2} \; ,\\
 w^2 & = 2 \frac{2 \, \lambda_h \, \mu_s^2 - \lambda_{hs} \, \mu_h^2}{4 \, \lambda_s \, \lambda_h - \lambda_{hs}^2} \; ,\;
\end{align}
and for the mass squared eigenvalues
\begin{equation}
 m_{1,2}^2 = \lambda_h \, v^2 + \lambda_s w^2 \mp \sqrt{(\lambda_s w^2 - \lambda_h v^2)^2 + \lambda_{hs}^2 w^2 v^2} \; .
\end{equation}
The corresponding mass eigenstates are
\begin{align}
 H_1 = s \sin \theta + h \cos \theta \nonumber \\
 H_2 = s \cos \theta - h \sin \theta 
\end{align}
with
\begin{equation}
 \tan 2\theta = \frac{\lambda_{hs} \, v \, w}{\lambda_h \, v^2 - \lambda_s \, w^2} \; .
\end{equation}
For small $\lambda_{hs}$ we find $m_1^2 \approx 2 \, \lambda_h \, v^2 \equiv m_h^2$ and $m_2^2 \approx 2 \, \lambda_s \, w^2 \equiv m_s^2$. This yields
\begin{equation}
 \theta \approx - \frac{\lambda_{hs} \, v \, w}{m_s^2 - m_h^2} + \mathcal{O}(\lambda_{hs}^3) \; .
\end{equation}

\end{appendix}

\providecommand{\href}[2]{#2}\begingroup\raggedright\endgroup

\end{document}